\documentclass[useAMS,usenatbib]{mn2e}
\usepackage{times}
\usepackage{amsmath}
\usepackage{amssymb}
\usepackage{longtable}
\usepackage{epsfig}

\title[NGC\,7009 spectrum]
{Very deep spectroscopy of the bright Saturn Nebula NGC\,7009 -- I. 
Observations and plasma diagnostics}

\author[X. Fang and X.-W. Liu]{X. Fang$^1$
       \thanks{E-mail: fangx@vega.bac.pku.edu.cn}\thanks{The complete line list Table~7 and Figure~16 are only available in electronic form at http://www.wiley.com} and X.-W. Liu$^{1,2}$\\
       $^1$Department of Astronomy, School of Physics, Peking University,
       Beijing 100871, P. R. China\\
       $^2$Kavli Institute for Astronomy and Astrophysics at Peking
       University, Beijing 100871, P. R. China}

\begin{document}

\date{Accepted . Received }

\pagerange{\pageref{firstpage}--\pageref{lastpage}} \pubyear{2010}

\maketitle

\label{firstpage}

\begin{abstract}
We present very deep CCD spectrum of the bright, medium-excitation planetary 
nebula NGC\,7009, with a wavelength coverage from 3040 to 11,000\,{\AA}. 
Traditional emission line identification is carried out to identify all the 
emission features in the spectra, based on the available laboratory atomic 
transition data. Since the spectra are of medium resolution, we use 
multi-Gaussian line profile fitting to deblend faint blended lines, most of 
which are optical recombination lines (ORLs) emitted by singly ionized ions of
abundant second-row elements such as C, N, O and Ne. Computer-aided 
emission-line identification, using the code {\sc emili} developed by Sharpee 
et al., is then employed to further identify all the emission lines thus 
obtained. In total about 1200 emission features are identified, with the 
faintest ones down to fluxes 10$^{-4}$ of H$\beta$. The flux errors for all 
emission lines, estimated from multi-Gaussian fitting, are presented. Plots of
the whole optical spectrum, identified emission lines labeled, are presented 
along with the results of multi-Gaussian fits

Of all the properly identified emission lines, permitted lines contribute 81 
per cent to the total line number. More than 200 O~{\sc ii} permitted lines 
are presented, as well as many others from N~{\sc ii} and Ne~{\sc ii}. Due to 
its relatively simple atomic structure, C~{\sc ii} presents few lines. Within 
the flux range 10$^{-2}$ -- 10$^{-4}$ H$\beta$ where most permitted lines of 
C~{\sc ii}, N~{\sc ii}, O~{\sc ii} and Ne~{\sc ii} fall, the average flux 
measurement uncertainties are about 10 to 20 per cent. Comparison is also made 
of the number of emission lines identified in the current work of NGC\,7009 
and those of several other planetary nebulae (PNe) that have been extensively 
studied in the recent literature, and it shows that our line-deblending 
procedure increases the total line number significantly, especially for 
emission lines with fluxes lower than 10$^{-3}$ of H$\beta$. Higher resolution
is still needed to obtain more reliable fluxes for those extremely faint 
emission lines, lines of fluxes of the order of 10$^{-5}$ -- 10$^{-6}$ of 
H$\beta$.

Plasma diagnostics using optical forbidden line ratios give an average electron
temperature of 10,020~K, which agrees well with previous results of the same 
object. The average electron density of NGC\,7009 derived from optical 
forbidden line ratios is 4290~cm$^{-3}$. The [O~{\sc iii}] 
$\lambda$4959/$\lambda$4363 nebular-to-auroral line ratio yields an electron
temperature of 9800~K. The ratio of the nebular continuum Balmer discontinuity
at 3646\,{\AA}\, to H~11 reveals an electron temperature of 6500~K, about 600~K
lower than the measurements published in the literature. The Balmer decrement 
reveals a density of about 3000~cm$^{-3}$. Also derived are electron 
temperatures from the He~{\sc i} line ratios, and a value of 5100~K from the
$\lambda$7281/$\lambda$6678 ratio is adopted. Utilizing the effective 
recombination coefficients newly available, we find an electron temperature 
around 1000~K from O~{\sc ii} ORL spectrum. Thus general pattern of electron 
temperatures, $T_\mathrm{e}$([O~{\sc iii}]) $\gtrsim$
$T_\mathrm{e}$(H~{\sc i} BJ) $\gtrsim$ $T_\mathrm{e}$(He~{\sc i}) $\gtrsim$ 
$T_\mathrm{e}$(O~{\sc ii}), which is seen in many PNe, is repeated in NGC\,7009.
Far-IR fine-structure lines, with observed fluxes adopted from the literature, 
are also used to derive $T_\mathrm{e}$ and $N_\mathrm{e}$. The [O~{\sc iii}] 
(52$\mu$m + 88$\mu$m)/$\lambda$4959 line ratio gives an electron temperature 
of 9260~K, and the 52$\mu$m/88$\mu$m ratio yields an electron density of 
1260~cm$^{-3}$.
\end{abstract}


\section{\label{introduction}
Introduction}

The Saturn Nebula NGC\,7009 is one of the best-known planetary nebulae (PNe), 
and has been extensively studied both observationally and theoretically. It is
a large, double-ringed, high-surface-brightness PN, with a pair of 
low-ionization knots ansae along its major axis. It has an H-rich O-type 
central star, with an effective temperature of 82,000K (M\'{e}ndez, Kudritzki 
\& Herrero \citealt{mendez1992}; Kingsburgh \& Barlow \citealt{kb1992}).

NGC\,7009 has been the subject of many investigations since the early twentieth
century. Berman \cite{berman1930} made the first photometric measurements and 
isophotic contours. Spectrophotometric measurements extending to the 
ultraviolet were carried out in late 1930's (Aller \citealt{aller1941}). Bowen
\& Wyse \cite{bw1939} and Wyse \cite{wyse1942} obtained spectra of NGC\,7009 
and estimated its chemical composition. In their work, nearly three hundred 
emission lines were detected, but only about 60 per cent were identified. 
Further studies were carried out by Aller \& Menzel \cite{am1945} and Aller 
\cite{aller1961}. NGC\,7009 is rich in emission lines, and is particularly well
known for its unusually rich and prominent O~{\sc ii} optical recombination 
lines ever since the early high-resolution photographic spectroscopy 
observations in the 1930's. Aller \& Kaler \cite{ak1964a} identified more than
100 O~{\sc ii} permitted transitions in the wavelength range 3100 -- 
4960\,{\AA}. In their longest-exposure spectrum of NGC\,7009, lines as weak as
0.02 on the scale where $I({\rm H}\beta)$ = 100 were detected. Kaler \& Aller 
\cite{ka1969} later reexamined the tracings of the long-exposure photographic 
plates used by Aller \& Kaler \cite{ak1964a} and \cite{ak1964b} and reported 
several dozen additional very faint lines just marginally above the plate 
noise level. Barker \cite{barker1983} obtained spectrophotometric observations 
at eight positions of NGC\,7009, covering a wide wavelength range from 
1400\,{\AA}\, to 10,000\,{\AA}. He found that the C$^{2+}$/H$^+$ abundance 
ratio derived from the C~{\sc ii} $\lambda$4267 optical recombination line 
(ORL) is significantly higher than that derived from the C~{\sc iii} 
$\lambda$\,$\lambda$1906,1909 collisionally excited UV intercombination lines,
a phenomenon first discovered by Perinotto \& Benvenuti \cite{pb1981}.

With the advent of modern high-quantum-efficiency and large-format linear 
detectors such as the IPCS and CCDs, more and more faint emission lines have 
been detected in the deep spectra of photoionized gaseous nebulae, including 
PNe and H~{\sc ii} regions. Albeit faint, many of them are of important 
diagnostic value to probe various nebular atomic processes such as radiative 
and dielectronic recombination, continuum and Bowen-like fluorescence and 
charge-exchange reactions. A very detailed study of the O~{\sc ii} optical 
permitted lines in the deep spectra of NGC\,7009 was presented by Liu et al. 
\cite{liu1995} who showed that for the 4f -- 3d transitions the departure from 
$LS$-coupling is important. They also found that the total elemental abundances
of C, N, and O relative to hydrogen based on the recombination line 
measurements are about a factor of 5 higher than the corresponding values 
derived from collisionally excited lines (CELs), a discrepancy previously known
to exist in the case of C$^{2+}$/H$^+$. A number of postulations have been 
proposed to explain this discrepancy, including temperature fluctuations and 
density inhomogeneities (Peimbert \citealt{peimbert1967}; Rubin 
\citealt{rubin1989}; Viegas \& Clegg \citealt{vc1994}), but failed to provide a
consistent interpretation of all observations. A bi-abundance model given by Liu et al. \cite{liu2000}, who postulate that nebulae contain H-deficient 
inclusions, provides a much better and natural explanation of the dichotomy. 
In this model, the optical recombination lines of heavy-element ions arise 
mainly from the ``cold" H-deficient component, while as the strong CELs are 
emitted predominantly from the warmer ambient plasma of `normal' chemical 
composition. Deep spectroscopic surveys (Tsamis et al. \citealt{tsamis2003}, 
\citealt{tsamis2004}; Liu et al. \citealt{liu2004a}, \citealt{liu2004b}; 
Wang et al. \citealt{wang2007}) and recombination line analysis of individual 
nebulae (Liu et al. \citealt{liu1995}; Liu et al. \citealt{liu2000}; Liu et 
al. \citealt{liu2001}; Liu et al. \citealt{liu2006a}) in the past decade has 
yielded strong evidence for the existence of such a ``cold" H-deficient 
component. Recent reviews on this topic are presented by Liu 
(\citealt{liu2003} and \citealt{liu2006b}).

In the study of emission line nebulae using faint heavy-element ORLs, which 
typically have intensities two to three magnitudes lower than H$\beta$, 
reliable line identifications become an important issue. Correct 
identifications of spectral lines are fundamental to all spectroscopic studies.
For lines commonly observed in astronomical spectra, a century of study has 
resulted in general agreement on transitions that give rise to strong lines 
observed at visible wavelengths. However, there is still much uncertainty 
about the proper identifications of many lines, particularly for fainter ones, 
and this problem is even more severe in other wavelength regions. As spectra 
approach fainter detection limits, the increasing number of features observed 
leads to a larger fraction of uncertain identifications. The effort involved 
in assigning correct and astronomically sound line identifications for the 
large numbers of emission lines detected in high signal-to-noise ratio (S/N) 
spectra can be daunting. Recent notable work on this topic includes 
P\'{e}quignot \& Baluteau \cite{pb1994}, Sharpee et al. \cite{sharpee2003} and
Zhang et al. \cite{zhang2005a}. In particular, Sharpee et al. 
\cite{sharpee2003} developed a computer-aided code to identify lines detected 
in emission line objects. The code automatically applies the same logic that 
is used in the traditional manual identification of spectral lines, working 
from a list of measured lines and a database of known transitions, and trying 
to find identifications based on the wavelengths and computed relative 
intensities of putative identifications, as well as on the presence of any 
other confirmed lines from the same multiplet or ion.

While robust emission line identification is a difficult task, especially for 
faint lines, today deep, high-resolution spectra of PNe (Liu et al. 
\citealt{liu1995}, \citealt{liu2000}; Sharpee et al. \citealt{sharpee2003}) 
and H~{\sc ii} regions (Esteban et al. \citealt{esteban1998}, 
\citealt{esteban1999}; Baldwin et al. \citealt{baldwin2000}) are routinely 
obtained. Since valuable information often results from the detection of 
previously unobserved low-abundant ionic species (e.g. P\'{e}quignot \& 
Baluteau \citealt{pb1994}), line identification remains a worthwhile 
investment. Emission line lists with robust identifications have been published
for a number of bright PNe and H~{\sc ii} regions, e.g. the Orion Nebula 
(3490 -- 7470\,{\AA}; Baldwin et al. \citealt{baldwin2000}), IC\,418 
(3510 -- 9840\,{\AA}; Sharpee et al. \citealt{sharpee2003}, 
\citealt{sharpee2004}), NGC\,7027 (3310 -- 9160\,{\AA}; Zhang et al. 
\citealt{zhang2005a}). In the current work, we present deep spectra and 
identified emission lines of NGC\,7009, another bright PN with archetypal rich 
and prominent heavy element ORLs. We illustrate the techniques used to deblend
faint lines, especially ORLs emitted by C$^+$, N$^+$ and O$^+$ ions. A detailed
analysis of those ORLs is the subject of a subsequent paper.

\section{\label{observe}
Observations and Data Reduction}

\subsection{\label{observe:part1}
Observations}

The spectra analyzed in the current work were observed from 1995 July to 2001 
June, using the ESO 1.52-m and the WHT 4.2-m telescopes. An observational 
journal is given in Table~\ref{observe_journal}. NGC\,7009 was observed in July
of 1995, 1996, 1999 and June of 2001, with the Boller \& Chivens long-slit 
spectrograph mounted on the ESO 1.52-m telescope. All spectra were secured 
with a long slit, whose width could be varied as shown in 
Table~\ref{observe_journal}. During the 1995 ESO 1.52-m run, the B\&C 
spectrograph was used with a Ford 2048$\times$2048 15$\mu$m$\times$15$\mu$m 
CCD. The slit was positioned at 79$^{\rm o}$, i.e. along the nebular major 
axis and passing through the two outlying ansae (c.f. the WFPC2 image of 
NGC\,7009 published by Balick et al. \citealt{balick1998}), and was offset 2--3
arcsec south of the central star (CS) in order to avoid the strong continuum 
emission from the CS, which, if included, would have had reduced the 
signal-to-noise ratios (S/N's) and the detectability of weak emission lines. 
The slit was about 3.5~arcmin long, and the slit width for all observations 
was 2~arcsec, except for one short exposure (60~s) for which an 8~arcsec wide 
slit was used (Table~\ref{observe_journal}). During the 1996 run, the CCD on 
the B\&C spectrograph was replaced by a thinned ultraviolet-enhanced Loral 
2048$\times$2048 15$\mu$m$\times$15$\mu$m chip of much improved quantum 
efficiency. For both observational runs, in order to reduce the read-out 
noise, the CCD was binned by a factor of 2 along the slit direction, yielding 
a spatial sampling of 1.63 arcsec per pixel projected on the sky. Several 
wavelength regions from the near ultraviolet (UV) atmospheric cut-off to 
approximately 5000\,{\AA}\, were observed using a 2400 line mm$^{-1}$ 
holographic grating, yielding a spectral resolution of approximately 
1.5\,{\AA}\, FWHM. Lower resolution spectra from 3520 to 7420\,{\AA}\, were also
obtained using a 600 groove mm$^{-1}$ grating. The $Hubble$ $Space$ $Telescope$
($HST$) standard stars, Feige~110 and (the nucleus of the PN) NGC~7293, were 
observed with an 8~arcsec wide slit for the purpose of flux calibration.

During the three nights' observations in 1999 at the ESO 1.52\,m, three 
wavelength regions were covered: 8105 -- 10,076\,{\AA}, 4697 -- 6724\,{\AA} 
and 3965 -- 4965\,{\AA}. The spectral resolution for the three sets of spectra
was about 3.0\,{\AA}\, FWHM. The third night was cloudy and no standard star 
was observed, so the spectra of the range 3965 -- 4965\,{\AA}\, were only 
wavelength calibrated. They are of limited usage given the low S/N's. Data 
from the first (8105 -- 10,076\,{\AA}) and second (4697 -- 6724\,{\AA}) nights 
were better. Data reduction was no easy task, especially for data obtained in 
the first night, very accurate wavelength calibration was needed in order to 
satisfactorily subtract the many bright sky OH emission lines present in the 
spectra. The first attempt to wavelength calibrate the spectra using arc lines
of an HeArFeNe lamp was not optimal as it was found that the geometric 
distortions of arc lines along the slit, which yielded crescent shaped line 
images with a curvature of about 1.2 pixels, were slightly different from 
those seen in the sky emission lines. This probably resulted from the fact 
that the light path of the comparison lamp was not exactly the same as that 
of the sky light. As a result, we opted to wavelength calibrate the spectra 
using the sky emission lines. The first attempt to wavelength calibrate a 
nebular spectrum using the sky emission lines detected in the same spectrum 
was not satisfactory, as for wavelengths longer than 9800\,\AA, no suitable 
sky emission lines not blended with nebular lines were available. Eventually, 
we calibrated all the first night's nebular spectra using the sky emission 
lines detected in a single narrow slit exposure of the standard star 
Feige~110. The second night's observations (4697 -- 6724\,{\AA}) were still 
wavelength calibrated using arc lines, as there were not enough number of sky 
lines in this wavelength region and the sky subtraction was less critical for 
this region. Despite all the efforts, subtraction of the many strong sky 
emission lines present in the 8105 -- 10,076\,{\AA}\, wavelength region was 
still not entirely satisfactory.

Observations at the WHT 4.2-m telescope were obtained using the ISIS double 
spectrograph during two observing runs in 1996 and 1997 
(Table~\ref{observe_journal}). For both the Blue and Red Arms, a Tek 
1024$\times$1024 24$\mu$m$\times$24$\mu$m chip was used, yielding a spatial 
sampling of 0.3576 arcsec per pixel projected on the sky. In 1996, gratings 
of 1200 and 600 groove mm$^{-1}$ were used in the Blue and Red Arms, 
respectively. The same set of gratings were used in 1997. Two wavelength 
regions, 3618 -- 4433\,{\AA}\, and 4190 -- 4989\,{\AA}, were covered by the 
Blue Arm and another two wavelength regions, 5176 -- 6708\,{\AA}\, and 6483 
-- 8005\,{\AA}, were covered by the Red Arm. During the 1996 run, the slit was
scanned across the whole nebula by uniformly driving the telescope 
differentially in Right Ascension (RA). The observations thus yielded average 
spectra for the whole nebula, which, when combined with the total H$\beta$ 
flux published in the literature and measured with a large entrance aperture, 
then yielded absolute fluxes of the whole nebula for all the emission lines 
detected in the spectra. The spectra obtained in 1997 were secured with a 
fixed slit oriented at PA = 90$^{\rm o}$ and passing through the CS. For both 
WHT runs, a 1~arcsec wide slit was used for nebular observations. Two $HST$ 
spectrophotometric standard stars, BD+28$^{\rm o}$ 4211 and HZ\,44, were 
observed using a 6~arcsec wide slit for the purpose of flux calibration.

There was still a gap in spectral coverage from 8000 to 8110\,{\AA}\, in the 
spectra described above. This gap was filled with 2001 observations at the 
1.52-m telescope, which covered a wide wavelength range from 7700 to 
11,100\,{\AA}\, (Table~\ref{observe_journal}). The spectral resolution was 
about 3.3\,{\AA}\, FWHM. Combined together, our observations of NGC\,7009 cover
the complete wavelength range from the near-UV 3000\,{\AA}\, to the near-IR 
11,100\,{\AA}.

\subsection{\label{observe:part2}
Data reduction}

\begin{figure*}
\begin{center}
\epsfig{file=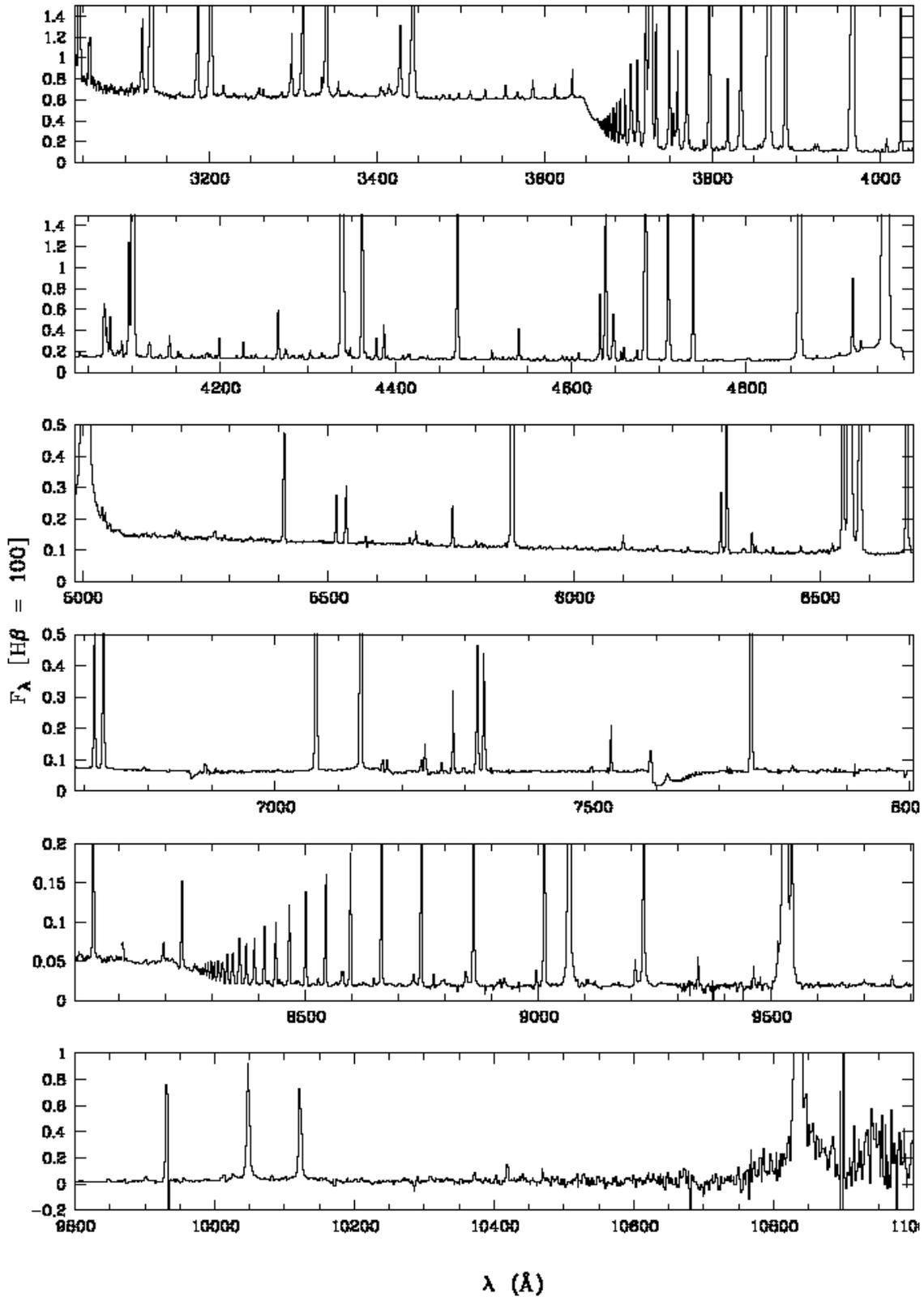,width=15cm,angle=0}
\caption{CCD spectra of NGC\,7009 from 3040 to 11,000\,{\AA}. Flux is 
normalized such that H$\beta$ = 100. Extinction has not been corrected for.}
\label{spectra}
\end{center}
\end{figure*}

All the spectra were reduced with standard procedures using 
{\sc midas}\footnote{{\sc midas} is developed and distributed by the European 
Southern Observatory.}. The spectra were bias-subtracted, flat-fielded, 
cosmic rays removed and wavelength calibrated using exposures of comparison 
lamps, and then flux-calibrated using wide slit observations of 
spectrophotometric standard stars. Ozone absorption bands that affect data 
points shortwards of 3400\,{\AA}\, were corrected for using observations of 
the standard stars Feige~110 and the CS of PN NGC~7293 taken with a 2~arcsec 
wide slit. As noted above, sky subtraction for spectra covering the wavelength 
range 8105 -- 10,076\,{\AA}\, in 1999 with the ESO 1.52-m telescope was not 
perfect. The extracted one-dimensional (1-D) spectrum of NGC\,7009 from 
3040 to 11,000\,{\AA}\, is shown in Fig.\,\ref{spectra}.

Detailed spectral processing to deblend and identify weak lines is illustrated 
in the following sections. Analysis of important ORLs that are detected or 
deblended in the spectrum is the topic of a subsequent paper.

\subsection{\label{observe:part3}
Reddening summary}

The logarithmic extinction at H$\beta$, $c$(H$\beta$), was derived by 
comparing the observed Balmer line ratios, H$\alpha$/H$\beta$ and 
H$\gamma$/H$\beta$, with the predicted Case~B values calculated by Storey \& 
Hummer \cite{sh1995} at $T_\mathrm{e}$ = $10,000$~K and $N_\mathrm{e}$ = 
$10,000$~cm$^{-3}$. This yielded a mean value of 0.174, larger than 0.07 given 
by Luo et al. \cite{luo2001} but close to 0.2 by Liu et al. \cite{liu1995}. As 
described in Luo et al. \cite{luo2001}, the discrepancy is probably partly 
caused by the different regions of the nebula being sampled, although the 
possibility that it is caused by the calibration uncertainties cannot be 
completely ruled out. We have dereddened the observed line fluxes by

\begin{equation}
  \label{deredden}
I(\lambda) = 10^{c(\rm{H}\beta) f(\lambda)} F(\lambda),
\end{equation}
where $f(\lambda)$ is the standard Galactic extinction curve for a
total-to-selective extinction ratio of $R$ = 3.1 (Howarth 
\citealt{howarth1983}), and $c({\rm H}\beta) = 0.174$.

\begin{table*}
\caption{Observational journal. All observations were taken with a 
2~arcsec wide slit unless otherwise specified.}
\label{observe_journal}
\begin{tabular}{cccccccc}
Date & Telescope & Wavelength & FWHM & Exposure Time & PA & Note\\
 & & Range~(\AA) & (\AA) & (sec) &  & \\
07/1995 & ESO 1.52~m & 3994--4983  &  1.50 & 2$\times$300,900,1418,5$\times$1800 & 79 & \\
        & ESO 1.52~m & 3523--7420  &  5.40 & 2$\times$30,60,60$^a$,1200 & 79 & \\
        & ESO 1.52~m & 3523--7420  &  5.40 & 60 & 79 & (1)\\
07/1996 & ESO 1.52~m & 3040--4048  &  1.50 & 5$\times$1800    & 79 & \\
07/1996 & ESO 1.52~m & 3994--4983  &  1.50 & 5,60,100,400,900 & 79 & \\
        & ESO 1.52~m & 3994--4983  &  1.50 & 2$\times$1800    & 79 & \\
07/1999 & ESO 1.52~m & 8105--10,076 &  3.00 & 4$\times$1200    & 79 & \\
07/1999 & ESO 1.52~m & 4697--6724  &  3.00 & 2$\times$1800,240,120 & 79 & \\
        & ESO 1.52~m & 4697--6724  &  3.00 & 60,1800 & 79 & \\
07/1999 & ESO 1.52~m & 3965--4965  &  3.00 & 2$\times$1200,600,120 & 79 & (2)\\
06/2001 & ESO 1.52~m & 3500--4805  &  1.50 & 1800          & 79 & \\
        & ESO 1.52~m & 7700--11,100 &  3.30 & 2$\times$1800 & 79 & \\
07/1996 & WHT 4.2~m  & 3618--4433  &  1.40 & 2$\times$600,2$\times$20 & Scanned & \\
07/1996 & WHT 4.2~m  & 4190--4989  &  1.40 & 2$\times$600,300,2$\times$20,2$\times$10 & Scanned & \\
07/1996 & WHT 4.2~m  & 5176--6708  &  2.70 & 2$\times$600,300,2$\times$20,2$\times$10 & Scanned & \\
08/1996 & WHT 4.2~m  & 6483--8005  &  2.90 & 2$\times$600,2$\times$20 & Scanned & \\
08/1997 & WHT 4.2~m  & 4104--4512  &  1.00 & 1200,431.79,120,30 & 79 & \\
08/1997 & WHT 4.2~m  & 4508--4918  &  1.00 & 1200,30   & 79 & \\
08/1997 & WHT 4.2~m  & 5166--5966  &  2.00 & 1200,120  & 79 & \\
08/1997 & WHT 4.2~m  & 5203--6006  &  2.00 & 428.19,30 & 79 & \\
08/1997 & WHT 4.2~m  & 6002--6809  &  2.00 & 1200,30   & 79 & \\
\end{tabular}
\begin{description}
\item [(1)] Observed with an 8~arcsec wide slit.
\item [(2)] Only wavelength calibrated spectra, no standards observed.
\end{description}
\end{table*}

\section{\label{identification}
Emission line identification}

Once the data have been reduced and high-quality 1-D spectra extracted, what 
follows is the most tedious and time-consuming process: emission line 
identifications. Three steps are involved in this procedure: (1) We first 
identify the strong (and obvious) emission lines manually, using the 
traditional empirical method, aided by available atomic transition data. Lists 
of identified emission lines in spectra of PNe published in the literature are 
also used. (2) We then use the technique of multi-Gaussian fitting to deblend 
the possible lines blended with a strong feature, except for those saturated 
ones that have no short-exposure, unsaturated data. This step requires 
strenuous efforts because our spectra are of medium resolution, and most, if 
not all, faint emission lines (mainly ORLs) are blended with adjacent stronger
features. Setting initial values for a multi-Gaussian fit is tricky, and can 
be more difficult when there are many, say more than ten, components in a 
single, broad feature. (3) After a complete emission line list has been 
generated after completing multi-Gaussian fitting across the whole spectrum, 
further identification is carried out with the computer-aided emission line 
identification code {\sc emili}, originally developed by Sharpee et al. 
\cite{sharpee2003}.

\subsection{\label{identification:empirical}
The traditional method}

The usual approach of identifying emission lines in high-quality spectra is to 
start with lists of line identifications available in the literature for 
spectra of objects of similar nature. It is generally necessary to manually 
work through various multiplet tables and line lists in order to arrive at an 
identification that makes physical sense in terms of the wavelength agreement 
with the laboratory value, anticipated intensity, and the presence or absence 
from the same multiplet or from the same ion. This process, which has been 
referred to as the traditional approach, is both tedious and prone to be 
incomplete (Sharpee et al. \citealt{sharpee2003}).

NGC\,7009 is an evolved medium-excitation PN exhibiting an extraordinary rich 
and prominent ORL spectrum of heavy element ions that is hardly rivaled by any 
other bright PNe of similar excitation class. Hitherto, fairly complete 
emission line identifications covering the whole optical wavelengths have been 
carried out for a few bright objects, including the high excitation class PN 
NGC\,7027 (Zhang et al. \citealt{zhang2005a}) and the low excitation class PN 
IC\,418 (Sharpee et al. \citealt{sharpee2003}; Sharpee et al. 
\citealt{sharpee2004}), amongst others. For NGC\,7027, a total of 1174 
identified emission lines have been tabulated, including 739 isolated features 
and more than two hundred blended ones without individual flux estimates. For 
IC\,418, a total of 807 emission lines are listed, including 624 with solid 
identifications and another 72 with possible identifications. While the 
physical conditions in those two PNe differ from those of NGC\,7009, we have 
found their line lists useful, in particular in identifying some features that 
are otherwise difficult to identify manually.

Liu et al. \cite{liu1995} identified and analyzed eight O~{\sc ii} ORL
multiplets (M1, M2, M5, M10, M12, M19, M20 and M26) belonging to the 3s -- 3p
and 3p -- 3d transitions, about 30 measurements belonging to the 3d -- 4f 
transition as well as eleven doubly excited (with O$^{2+}$ parentage other 
than $^3$P) ORLs in the blue optical spectrum of NGC\,7009. In the same 
object, Luo et al. \cite{luo2001} reported and analyzed several dozen 
Ne~{\sc ii} ORLs belonging to transitions 3s -- 3p, 3p -- 3d and 3d -- 4f. 
One expects that most lines emitted by NGC\,7009 are of intermediate excitation
energies, and second-row elements (C, N, O and Ne) are mostly doubly ionized. 
Triply ionized species should exist to some extent, whereas those of even 
higher ionization stages must be negligible.

We use {\sc midas} to process the continuum-subtracted 1-D spectra. The basic 
observational information, such as central wavelengths, FWHMs and fluxes 
(normalized to a scale where H$\beta$ = 100), of all the obvious, isolated 
emission features in the spectra is obtained using {\sc midas}. For those 
features with line profiles that obviously deviate from Gaussian, e.g., 
features that suffer from serious blending, the peak wavelength is adopted as 
the central wavelength. S/N ratio for each feature is obtained by measuring 
the standard deviation of the local continuum near the emission feature. All 
the observed wavelengths are corrected for the Doppler shifts (NGC\,7009 is 
blue-shifted by about $100$~km/s) estimated from the hydrogen Balmer and 
Paschen lines.

We manually identify all the isolated emission features by comparing the 
measured central wavelengths with laboratory values available from the atomic 
spectral line lists compiled by Hirata \& Horaguchi \cite{hh1995}, after 
taking into account the Doppler shifts. The line lists of Hirata \& Horaguchi 
\cite{hh1995} include only dipole transitions. For forbidden transitions, 
emission line lists of other PNe (e.g. NGC\,7027, IC\,418) from the literature 
are used. For features for which we cannot find reasonable identifications in 
the lists of either Hirata \& Horaguchi \cite{hh1995} or from the literature, 
online atomic transition database\footnote{Atomic Line List v2.05 by Dr. Peter
van Hoof, website: http://www.pa.uky.edu/~peter/newpage/ .} is used as an aid. 
This step of manual identification is based on wavelength matching only, and 
could be unreliable for some features. The measured fluxes are not used 
because at this stage many of lines suffer from line-blending issues. Once 
this first round of emission-line identification is complete, we scan through 
the list checking for the presence of the other components belonging to the 
same multiplet if one of the components is identified. About 700 isolated 
emission features are obtained in this preliminary line list.

\subsection{\label{identification:enlargement}
Enlargement of the emission line list}

The preliminary line list created above is incomplete. Since the spectra are 
of medium resolution, $1.5$\,{\AA}\, at short wavelengths and $3.0$\,{\AA}\, 
at longer ones, many faint ORLs are partially blended with or even entirely 
embedded in strong features. We check for lines that should be present but are
blended and add them to the emission line list. This empirical approach should 
proceed with great care. We search through available atomic database, 
including high resolution spectra of PNe in the literature. Although the 
physical conditions in different PNe are different, a large number of ORLs are 
commonly observed, as judged from a comparison of line lists of different 
objects. In addition to the most recent atomic database three criteria are 
used to decide whether a line should be present in the spectrum and thus be 
included in the line list:

(1) Elemental abundances: The most abundant heavy elements are O, N, C and Ne 
from the second row of the table of chemical elements, thus we consider the 
presence of their ORLs prior to those from other less abundant elements.

(2) Components within the same multiplets: If one or several components of a 
given multiplet have been detected with solid identifications, we assume that 
the other components must also be present. Only components that are too faint 
to be of any significance, say more than two orders of magnitude fainter than 
the principal component, are neglected.

(3) Ionization potential and excitation energy: In cases where not even a 
single component of a given multiplet is seen, but this multiplet probably 
should exist as judged from its excitation energy as well as ionization 
potential of the emitting ion, we add the multiplet to the line list. Only 
components that may be of any significance are included.

The emission line list enlarged in this way may still be incomplete. Limited 
by the spectral resolutions and S/N's, reliable fluxes for many faint blended 
lines are difficult to obtain. However, the enlarged emission line list, now 
containing more than 1300 transitions, has nearly doubled the size of the 
original preliminary one.

\subsection{\label{identification:fit}
Multi-Gaussian fits}

Many of those faint ORLs are of great value for astrophysical diagnostics. 
In order to obtain reliable fluxes for them, deblending is often needed. 
Assuming all lines have a Gaussian profile, we use multi-Gaussian fitting to 
deblend the lines using procedures in {\sc midas}. The fitting proceeds from 
blue to red wavelengths and covers the whole wavelength range 
(3040 -- 11,000\,{\AA}) of the continuum-subtracted 1-D spectra, with each fit 
covering a spectral segment of about 20--30\,{\AA}\, at short wavelengths 
(FWHM~$\sim$~1.5\,{\AA}) and about 50--60\,{\AA}\, at long wavelengths 
(FWHM~$\sim$~3.0\,{\AA}). Much wider spectral segments are not favored 
because: (1) A wide spectral range often contains many emission components. 
Fitting the whole range in a single go will involve a large number of 
free parameters and thus may result in large uncertainties in the output. (2) 
The maximum number of components that {\sc midas} can handle in a single 
multi-Gaussian fit is 16. The fit may also diverge if too many components 
crowd in a narrow wavelength range with adjacent lines of wavelength 
difference $\Delta\lambda<$~0.5\,{\AA}.

For each segment fitted, we make sure that its blue and red ending points are 
equal or close to the local continuum level as judged by eyes. If several 
broad features are partially blended with each other and cover a wide spectral
range, we split them into two smaller ones and the split point is chosen where 
the separation of the features is the largest. This is also judged by eyes.

For each multi-Gaussian fit to a spectral segment we first set the initial 
values for the parameters of each Gaussian component -- height (CCD pixel 
value), central wavelength and width (i.e. FWHM). In general, all Gaussian 
components considered in a given fit are assumed to have the same line width, 
and the wavelength differences between individual components are fixed to 
their known laboratory values. Even if the initial values have been set with 
care, several tries are often needed before the program converges and 
reasonable fitting results are obtained. Fig.\,\ref{4625-4680} shows the 
Gaussian profile fits to the spectral range 4625 -- 4680\,{\AA}, where the 
O~{\sc ii} M1 lines locate. Some specific cases are noted below:

\begin{figure*}
\begin{center}
\epsfig{file=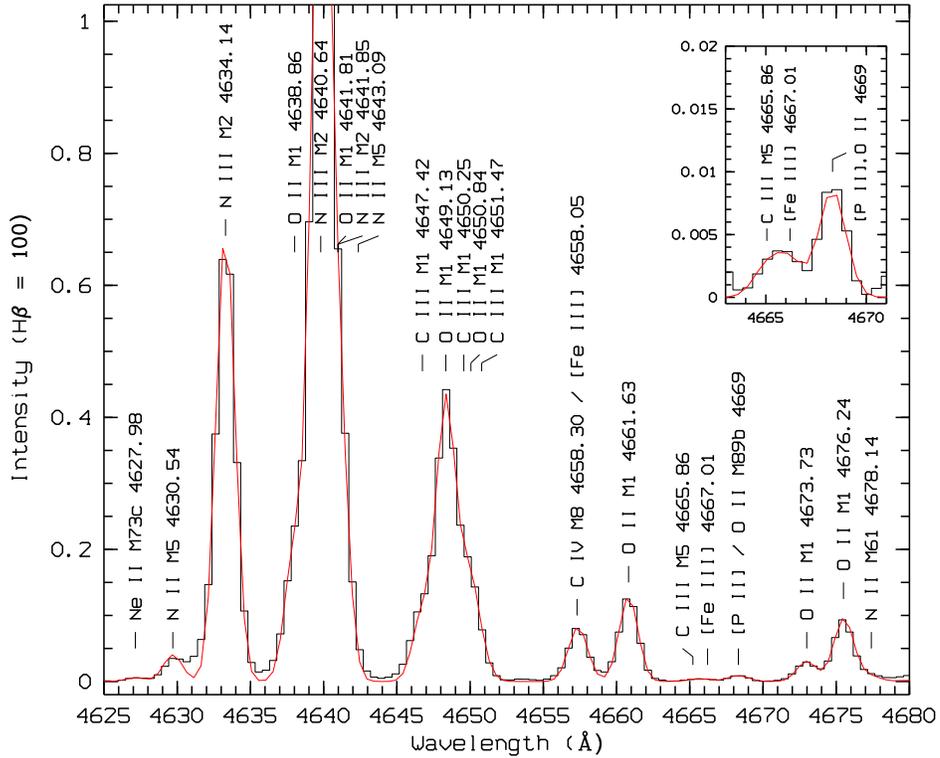,width=10cm,angle=-90}
\caption{Spectrum of NGC\,7009 from 4625 to 4680\,{\AA} showing the O~{\sc ii}
M1 lines and other blended emission features. The continuous curve is the sum 
of Gaussian profile fits. The inset zooms in the weak features of 
[Fe~{\sc iii}] $\lambda$4667.01, which is blended with C~{\sc iii} M5 
3p$^{\prime}$~$^3$P$_{2}$ -- 3s$^{\prime}$~$^3$P$^{\rm o}_{2}$ 
$\lambda$4665.86, and O~{\sc ii} M89b 4f~D[2]$^{\rm o}_{5/2}$ -- 
3d~$^2$D$_{3/2}$ $\lambda$4669.27, which is blended with [P~{\sc ii}] 
3p$^2$~$^1$S$_{0}$ -- 3p$^2$~$^3$P$_{1}$ $\lambda$4669.25 and contributes 
about 40 per cent to the total flux of the blend at $\lambda$4669. Continuum 
has been subtracted and fluxes normalized such that H$\beta$ = 100. Extinction
has not been corrected for.}
\label{4625-4680}
\end{center}
\end{figure*}

(1) For a blended feature, the wavelength of the strongest component can be 
easily estimated. For the other fainter blended components, we fix their 
wavelengths relative to those stronger ones, utilizing their known laboratory 
wavelengths. All components are assumed to have the same FWHM. There are cases 
where the strongest emission component obviously deviates from a Gaussian 
profile. In such case we set a large FWHM to it while keeping the Gaussian 
assumption.

(2) For cases where several faint lines blend with a strong one with close 
wavelengths, say $\Delta\lambda$~$<$~0.5\,{\AA}, accurate fluxes are almost 
impossible to obtain even with multi-Gaussian fitting. In such cases fluxes of 
fainter lines are estimated from the ionic abundances (often deduced from lines
free of serious blending) using available atomic data, i.e. the effective 
recombination coefficients for recombination lines, and collisional strengths 
for forbidden lines. If a faint line contributes little, say less than 5 per 
cent, to the total intensity of the strong feature, we assume that the faint 
line is negligible. 

\begin{figure}
\begin{center}
\epsfig{file=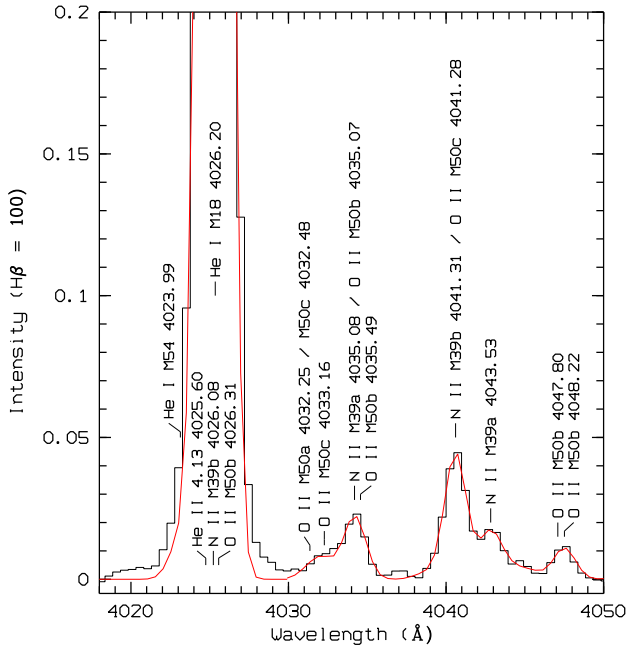,width=8.5cm,angle=-90}
\caption{Spectrum of NGC\,7009 from 4018 to 4050\,{\AA} showing the emission 
lines identified in this wavelength range. Here the strongest emission feature
is the He~{\sc i} M18 $\lambda$4026.20 line, with at least four weak lines 
blended with it: He~{\sc ii} 4.13 $\lambda$4025.60, N~{\sc ii} M39b 
$\lambda$4026.08, O~{\sc ii} M50b $\lambda$4026.31 and He~{\sc i} M54 
$\lambda$4023.99. Other emission lines present in this wavelength range, 
mostly N~{\sc ii} and O~{\sc ii} permitted lines, are also marked. The 
continuous curve represents the sum of multi-Gaussian fits. The spectrum has 
been continuum subtracted and flux normalized such that H$\beta$ = 100. 
Extinction has not been corrected for.}
\label{4018-4050}
\end{center}
\end{figure}

Fig.\,\ref{4018-4050} is an example illustrating the case of deblending the 
He~{\sc i} $\lambda$4026 feature. Here the N~{\sc ii} M39b 4f~2[5]$_{4}$ -- 
3d~$^3$F$^{\rm o}_{3}$ $\lambda$4026.08 and O~{\sc ii} M50b 
4f~F[3]$^{\rm o}_{5/2}$ -- 3d~$^4$F$_{3/2}$ $\lambda$4026.31 lines are blended 
with the stronger He~{\sc i} M18 5d~$^3$D -- 2p~$^3$P$^{\rm o}$ 
$\lambda$4026.20 line, with wavelength differences less than 0.5\,{\AA}. Also 
embedded in this feature are the He~{\sc ii} 4.13 13g~$^2$G -- 
4f~$^2$F$^{\rm o}$ $\lambda$4025.60 and He~{\sc i} M54 7s~$^1$S$_{0}$ -- 
2p~$^1$P$^{\rm o}_{1}$ $\lambda$4023.99 lines.

Fluxes of the blended N~{\sc ii} $\lambda$4026.08 and O~{\sc ii} 
$\lambda$4026.31 lines are estimated from the N~{\sc ii} and O~{\sc ii} 
effective recombination coefficients calculated by Fang et al. \cite{fang2010}
and Storey (private communications), 
respectively, using the equation,

\begin{equation}
\label{flux}
\frac{I\left(\lambda\right)}{I\left({\rm H}\beta\right)} = 
\frac{\alpha_{\rm eff}\left(\lambda\right)}{\alpha_{\rm eff}\left({\rm H}\beta\right)}\frac{\lambda}{4861} \frac{{\rm X}^{+}}{{\rm H}^{+}},
\end{equation}
where $\alpha_{\rm eff}(\lambda)$ and $\alpha_{\rm eff}({\rm H}\beta)$ are the 
effective recombination coefficients for the emission line $\lambda$ and 
H$\beta$ respectively, and X$^+$/H$^+$ the ionic abundance ratio (N$^{2+}$ for 
the N~{\sc ii} line and O$^{2+}$ for the O~{\sc ii} line). N$^{2+}$/H$^+$ and 
O$^{2+}$/H$^+$ abundance ratios are adopted from Liu et al. \cite{liu1995}. 
Here we assume $T_\mathrm{e}$ = 1000~K, as diagnosed from N~{\sc ii} and 
O~{\sc ii} recombination lines, and $N_\mathrm{e}$ = 4300~cm$^{-3}$, the 
average electron density of NGC\,7009 deduced from optical CEL ratios 
(Section~4). Details for the derivations of $T_\mathrm{e}$ based on ORLs will 
be presented in a subsequent paper. The fluxes of the blended N~{\sc ii} 
$\lambda$4026.08 and O~{\sc ii} $\lambda$4026.31 lines thus estimated 
contribute, respectively, 0.74 and 0.09 per cent to the He~{\sc i} 
$\lambda$4026 feature.

For the He~{\sc ii} $\lambda$4025.60 line, its flux is calculated from the 
He$^{++}$/H$^{+}$ abundance ratio using the hydrogenic recombination theory of 
Storey \& Hummer \cite{sh1995}, with $T_\mathrm{e}$ and $N_\mathrm{e}$ assumed 
to be 10,000~K and 10,000~cm$^{-3}$, respectively. The estimated He~{\sc ii} 
line flux contributes 5.7 per cent to the He~{\sc i} $\lambda$4026 feature. The 
contribution of the He~{\sc i} $\lambda$4023.99 line is estimated from the 
Case~B predictions of Brocklehurst \cite{brocklehurst1972}, and it amounts 
about 0.66 per cent. In the latter case, we assume $T_\mathrm{e}$ = 5000~K, as 
is derived from the He~{\sc i} $\lambda$7281/$\lambda$6678 line ratio, and 
$N_\mathrm{e}$ = 10,000~cm$^{-3}$.

\begin{figure}
\begin{center}
\epsfig{file=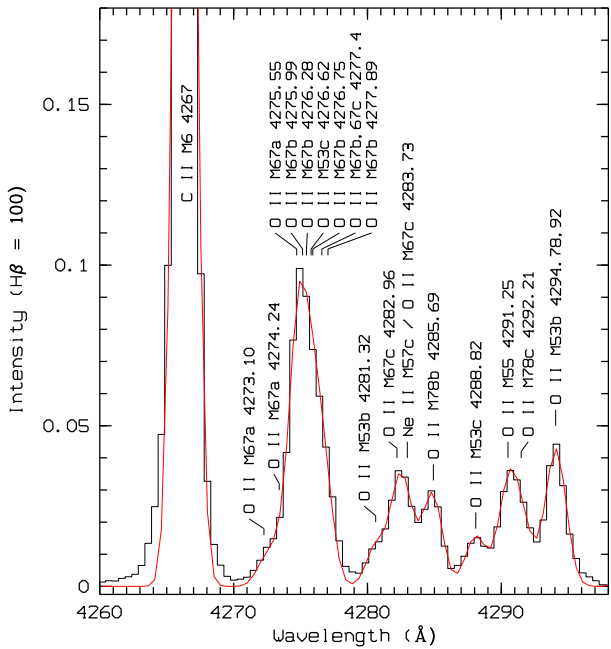,width=8.5cm,angle=-90}
\caption{Spectrum of NGC\,7009 from 4260 to 4298\,{\AA}. The very broad 
emission feature at 4275\,{\AA}\, which is formed by more than ten O~{\sc ii} 
ORLs from the 4f -- 3d transition array blended together, is clearly seen, 
with positions and wavelengths of the components labeled.}
\label{4260-4298}
\end{center}
\end{figure}

(3) There are cases where several faint ORLs blend together and their 
wavelengths do not differ much. Such feature often exhibits an irregular 
spectral shape, where only the relatively stronger ones can be easily 
discerned from the obvious peaks. In order to estimate fluxes for the fainter 
ones, we resort to theoretical atomic data. If the components belong to the 
same multiplet, their relative intensities are set to equal to the predicted 
values, i.e., $I(\lambda_1)$ : $I(\lambda_2)$ : ... = 
$\alpha_{\rm eff}(\lambda_1)$/$\lambda_1$ : 
$\alpha_{\rm eff}(\lambda_2)$/$\lambda_2$ : ..., 
where $\alpha_{\rm eff}(\lambda_{i})$ is the effective recombination 
coefficient of the blended component $i$, and $I(\lambda_i)$ its intensity. 
The effective recombination coefficients adopted here are calculated assuming 
appropriate physical conditions ($T_\mathrm{e}$ and $N_\mathrm{e}$) under 
which ORLs are emitted. For NGC\,7009, we assume $T_\mathrm{e}$ = $1000$~K and
$N_\mathrm{e}$ = 4300~cm$^{-3}$. Here 4300~cm$^{-3}$ is the average electron 
density deduced in Section~4. Components predicted to have intensities 
negligible compared to others, are ignored. Fig.\,\ref{4260-4298} shows a very
broad feature between the wavelength range 4270 -- 4280\,{\AA} formed by more 
than ten O~{\sc ii} lines from the 4f -- 3d transition array blended together.
Within this broad feature, the strongest component is M67a 
4f~F[4]$^{\rm o}_{9/2}$ -- 3d~$^4$D$_{7/2}$ $\lambda$4275.55 with a fitted 
intensity of 0.11 (on a scale where H$\beta$ = 100). Fluxes of other 
components are estimated from the O~{\sc ii} effective recombination 
coefficients of Storey (2008), and the values are 10$^{-4}$ of H$\beta$ flux 
or lower.

The above cases illustrate the procedure that we deblend the spectra of 
NGC\,7009 and estimate the contributions of the blended faint lines whose 
reliable measurements are difficult. We obtain an enlarged line list 
containing 1440 individual transitions, 169 of which are estimated from the 
available atomic data and thus have large uncertainties in their fluxes. The 
sum of all the multi-Gaussian fits is overplotted on the observed spectrum in 
Fig.\,16 for comparison.

\subsection{\label{identification:emili}
Finalization of the line list}

Up to this stage, the emission line list remains preliminary, even though a 
large number of faint lines have been added. The identifications are done 
manually and may contain spurious ones, especially for those `isolated' 
features without any other corroborative components from the same multiplet. 
Care must also be exercised for those that appear only in the line list of
NGC\,7009. Further more, the atomic database used could be incomplete, or 
there are more than one candidate transitions for the observed feature. A more
systematic approach to line identification is needed to generate reasonable 
results for all emission features that are either detected or deblended
properly, if necessary. Here the emission line identification code, 
{\sc emili}\footnote{{\sc emili} is developed by Dr. Brian Sharpee et al. and 
is designed to aid in the identification of weak emission lines, particularly 
the weak recombination lines seen in high dispersion and high signal-to-noise 
spectra.}, has been used to aid and clarify identifications of faint emission 
lines. Technic details of {\sc emili} and its operation are documented in 
Sharpee et al. \cite{sharpee2003}. Previously mentioned in 
Section~\ref{identification:fit}, the intensities of a number of blended faint
lines (mainly ORLs) are simply estimated from the atomic data (e.g. effective 
recombination coefficients) at assumed $T_\mathrm{e}$'s and $N_\mathrm{e}$'s 
and these estimated intensities could be of large uncertainties. These lines 
are excluded from the input list of the {\sc emili} code, and henceforth 
referred to simply as estimated lines.

{\sc emili} reads in the basic information (wavelengths and errors, line widths
in velocity, and line intensities, etc.) of the 1271 emission transitions in 
the line list we have constructed, searches for all the possible candidates in 
the atomic transition database for each of those input transitions and decides 
which candidate matches the observation best. {\sc emili} automatically applies
the same logic that is used in the traditional manual identification of 
spectral lines, working on a list of measured lines and a database of known 
transitions, and trying to find identifications based on wavelength agreement 
and relative computed intensities of putative identifications, and on the 
presence of other confirming lines from the same multiplet or ion.

An analysis of the code output shows: (1) For strong lines, e.g., most CELs,
hydrogen lines, relatively strong helium lines, and some relatively strong 
ORLs emitted by C, N, O and Ne ions, the code output agrees well with the 
empirical identifications obtained manually. (2) For most faint ORLs emitted 
by C, N, O and Ne ions as well as by ions of the third-row elements (e.g., 
Mg~{\sc ii}, Si~{\sc ii}, Si~{\sc iii}, etc.) identifications returned by the
code generally agree with empirical identifications, except for a few faint 
lines with fluxes $\simeq$~10$^{-4}$ to 10$^{-5}$ times the H$\beta$. For this
small number of lines, {\sc emili} gives identifications different from the 
original assignments, and this adds ambiguity to their identifications. For 
these lines, we search the whole line list and choose the candidate that have 
highest number of solidly identified lines from the same multiplet or from the
same ion. (3) A few dozen emission lines are unidentified. We attribute this 
to either the lack of relevant atomic data, or large errors from the 
multi-Gaussian fits, or spurious features. For transitions that have 
alternative identifications given by {\sc emili}, we list all the candidates 
and mark them as blends (the fourth column of Table~\ref{linelist}). In total,
there are 1170 identified transitions in the final emission line list plus 28
unknown features, 235 alternative identifications given by {\sc emili} and 
73 questionable ones. Most of the latter ones are identified as Ne~{\sc i}, 
Fe~{\sc i}, Co~{\sc i}, Ni~{\sc i} and etc. The identifications as well as 
the observed and dereddened line fluxes (on a scale where H$\beta$ is 100) 
are presented in Table~\ref{linelist}. The 169 blended faint lines, whose 
intensities are simply estimated from atomic data, are also included in the 
line list, and their flux errors (the last column of Table~\ref{linelist}) 
could be large and are labeled by ``:". In Fig.\,16, 
identifications of all lines present in the spectrum are marked.

\section{\label{plasma_diagnostics}
Plasma diagnostics}

Electron densities and temperatures are deduced from optical CEL ratios of 
heavy element ions. Also derived are temperatures from the hydrogen continuum
Balmer and Paschen discontinuities, and from the He~{\sc i} and He~{\sc ii} 
recombination spectrum, the $N_\mathrm{e}$'s from the Balmer and Paschen 
decrements. The enhancement of the [N~{\sc ii}], [O~{\sc ii}] and [O~{\sc iii}]
auroral lines contributed by recombination excitation is discussed, and 
estimation of the enhancement is provided.

\subsection{\label{plasma_diagnostics:part1}
$T_\mathrm{e}$'s and $N_\mathrm{e}$'s from CELs}

\begin{figure}
\begin{center}
\epsfig{file=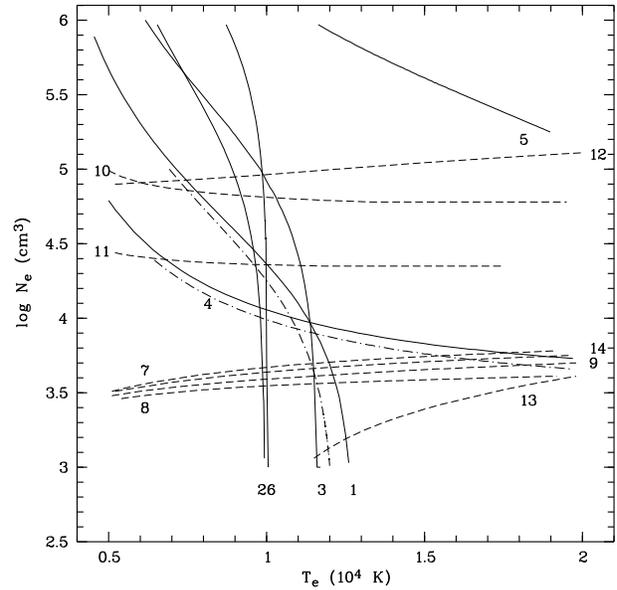,width=8cm,angle=0}
\caption{Plasma diagnostic diagram. Each curve is labeled with an ID number 
given in Table~\ref{diagnostics}. The solid curves are temperature-sensitive 
diagnostics, and the dashed curves are density-sensitive diagnostics. The 
dash-dot curves represent temperature diagnostics of [N~{\sc ii}] 
(labeled as 1) and [O~{\sc ii}] (labeled as 4), after the contributions by 
recombination excitation to the auroral lines have been corrected for, whereas
the solid curves represent those without corrections.}
\label{diagram}
\end{center}
\end{figure}

The spectrum of NGC\,7009 reveals many CELs, useful for nebular electron 
density and temperature diagnostics and abundance determinations. Adopting the
atomic data from the references given in Table~\ref{references:cel} and 
solving the level populations for multilevel ($\ge~5$) atomic models, we have 
determined electron temperatures and densities from a variety of CEL ratios. 
The results are listed in Table~\ref{diagnostics}. Electron temperatures were 
derived assuming a constant electron density of $\log$~$N_\mathrm{e}$ = 
3.633~(cm$^{-3}$), or 4290~cm$^{-3}$, the average value yielded by the four 
density-sensitive CEL diagnostic ratios, [O~{\sc ii}] 
$\lambda$3726/$\lambda$3729, [S~{\sc ii}] $\lambda$6731/$\lambda$6716, 
[Ar~{\sc iv}] $\lambda$4740/$\lambda$4711 and [Cl~{\sc iii}] 
$\lambda$5537/$\lambda$5517. The four [Fe~{\sc iii}] line ratios, 
$\lambda$4733/$\lambda$4754, $\lambda$4701/$\lambda$4733, 
$\lambda$4881/$\lambda$4701 and $\lambda$4733/$\lambda$4008, yield much higher
electron densities for $T_\mathrm{e}$ = 10,020~K, the average temperature 
deduced from a number of optical CEL ratios. We have thus ignored densities 
yielded by the [Fe~{\sc iii}] line ratios in calculating the average electron
density. A plasma diagnostic diagram based on CEL ratios is plotted in 
Fig.\,\ref{diagram}. The temperature diagnostic ratio, [O~{\sc i}] 
($\lambda$6300 + $\lambda$6363)/$\lambda$5577 is not used because of poor sky
subtraction for the auroral line $\lambda$5577, making its intensity quite 
unreliable.

The [O~{\sc ii}] nebular-to-auroral line ratio ($\lambda$7320 + 
$\lambda$7330)/$\lambda$3729 yields an electron temperature 
of about 19,000~K, which is abnormally high, if one assumes an electron 
density of Log$N_\mathrm{e}$ = 3.674 (Table~\ref{diagnostics}), as yielded 
by the [O~{\sc ii}] $\lambda$3726/$\lambda$3729 line ratio. Here we assume 
$N_\mathrm{e}$ = 10,000~cm$^{-3}$ for [O~{\sc ii}], and then the [O~{\sc ii}] 
nebular-to-auroral line ratio yields a temperature of 9850~K, which is quite 
reasonable.

\begin{table}
\caption{References for CEL atomic data.}
\label{references:cel}
\begin{tabular}{lll}
\hline
ion & Transition probabilities & Collision strengths\\
\hline
N~{\sc ii}   & Nussbaumer \& Rusca \citealt{nr1979}   & Stafford et al. \citealt{stafford1994}\\
O~{\sc i}    & Baluja \& Zeippen \citealt{bz1988}     & Berrington \citealt{berrington1988}\\
O~{\sc ii}   & Zeippen \citealt{zeippen1982}          & Pradhan \citealt{pradhan1976}\\
O~{\sc iii}  & Nussbaumer \& Storey \citealt{ns1981}  & Aggarwal \citealt{aggarwal1983}\\
Ne~{\sc iii} & Mendoza \citealt{mendoza1983}          & Butler \& Zeippen \citealt{bz1994}\\
S~{\sc ii}   & Mendoza \& Zeippen \citealt{mz1982b}   & Keenan et al. \citealt{keenan1996}\\
             & Keenan et al. \citealt{keenan1993}     & \\
S~{\sc iii}  & Mendoza \& Zeippen \citealt{mz1982a}   & Mendoza \citealt{mendoza1983}\\
Cl~{\sc iii} & Mendoza \citealt{mendoza1983}          & Mendoza \citealt{mendoza1983}\\
Ar~{\sc iii} & Mendoza \& Zeippen \citealt{mz1983}    & Johnson \& Kingston \citealt{jk1990}\\
Ar~{\sc iv}  & Mendoza \& Zeippen \citealt{mz1982b}   & Zeippen et al. \citealt{zeippen1987}\\
Fe~{\sc iii} & Nahar \& Pradhan \citealt{np1996}      & Zhang \citealt{zhang1996}\\
\hline
\end{tabular}
\end{table}

\begin{table}
\centering
\caption{Plasma diagnostics.}
\label{diagnostics}
\begin{tabular}{cll}
\cline{1-3}\\
ID & Diagnostic & Result\\
\hline
 & & $T_\mathrm{e}$~[K]\\
1 & [N~{\sc ii}] ($\lambda$6548 + $\lambda$6584)/$\lambda$5754 & 10,780$^a$\\
2 & [O~{\sc iii}] ($\lambda$4959 + $\lambda$5007)/$\lambda$4363 & 10,940\\
3 & [S~{\sc iii}] ($\lambda$9531 + $\lambda$9069)/$\lambda$6312 & 11,500\\
4 & [O~{\sc ii}] ($\lambda$7320 + $\lambda$7330)/$\lambda$3729 & 9850$^b$\\
5 & [O~{\sc i}] ($\lambda$6300 + $\lambda$6363)/$\lambda$5577 & --$^c$\\
6 & [Ar~{\sc iii}] $\lambda$7135/$\lambda$5192 & 10,050\\
 & [O~{\sc iii}] $\lambda$4959/$\lambda$4363 & 9810\\
 & [O~{\sc iii}] (52$\mu$m + 88$\mu$m)/$\lambda$4959 & 9260\\
 & [Ne~{\sc iii}] (15.5$\mu$m + 36$\mu$m)/($\lambda$3868 + $\lambda$3967) & 9010\\
 & [S~{\sc ii}] ($\lambda$6717 + $\lambda$6731)/($\lambda$4069 + $\lambda$4076) & 9770\\
 & Average optical CEL temperature & 10,020$^d$\\
 & He~{\sc i} $\lambda$7281/$\lambda$5876 & 3850\\
 & He~{\sc i} $\lambda$7281/$\lambda$6678 & 5100\\
 & He~{\sc i} $\lambda$5876/$\lambda$4471 & 4360\\
 & He~{\sc i} $\lambda$6678/$\lambda$4471 & 9690\\
 & H~{\sc i} BJ / H11 & 6420\\
 & H~{\sc i} PJ / P11 & 6750$\pm$160\\
 & He~{\sc i} Jump at 3421\,{\AA} & 7800$\pm$200\\
 & He~{\sc ii} Jump at 5694\,{\AA} & 11,000$\pm$2000\\
\\
 & & $N_\mathrm{e}$~[cm$^{-3}$]\\
7 & [Ar~{\sc iv}] $\lambda$4740/$\lambda$4711 & 4890\\
8 & [Cl~{\sc iii}] $\lambda$5537/$\lambda$5517 & 3600\\
9 & [S~{\sc ii}] $\lambda$6731/$\lambda$6716 & 4100\\
10 & [Fe~{\sc iii}] $\lambda$4733/$\lambda$4754 & 64,560\\
11 & [Fe~{\sc iii}] $\lambda$4701/$\lambda$4733 & 22,860\\
12 & [Fe~{\sc iii}] $\lambda$4881/$\lambda$4701 & 93,330\\
13 & [Fe~{\sc iii}] $\lambda$4733/$\lambda$4008 & --$^e$\\
14 & [O~{\sc ii}] $\lambda$3726/$\lambda$3729 & 4720\\
 & Adopted optical CEL density & 4290\\
 & [O~{\sc iii}] 52$\mu$m/88$\mu$m & 1260\\
 & [Ne~{\sc iii}] 15.5$\mu$m/36$\mu$m & 11,480\\
 & H~{\sc i} Balmer decrement & 3000\\
 & H~{\sc i} Paschen decrement & 1000$\sim$3000\\
\hline
\end{tabular}
\begin{description}
\item [$^a$] Probably unreliable. Here an electron density Log$N_\mathrm{e}$ 
= 4.0 is assumed. The intensity of the auroral line $\lambda$5754 has been 
corrected for the contribution from recombination excitation 
(c.f. Section~4.3), which amounts to 9.5 per cent.
\item [$^b$] Assuming $LogN_\mathrm{e}$ = 4.0. The intensity of the two 
auroral lines $\lambda\lambda$7320, 7330 have been corrected for the 
contribution by recombination excitation (c.f. Section~4.3), which is about 13
per cent.
\item [$^c$] The intensity of the $\lambda$5577 line is unreliable due to poor
sky subtraction.
\item [$^d$] Average temperature derived from the [O~{\sc ii}], [O~{\sc iii}], 
[S~{\sc iii}] and [Ar~{\sc iii}] nebular-to-auroral line ratios.
\item [$^e$] The loci of this diagnostic delineates very high temperatures, 
from 11,500 to 19,800~K for the density range 
3.06~$\leq$~Log$N_\mathrm{e}$~$\leq$~3.61.
\end{description}
\end{table}

\subsection{\label{plasma_diagnostics:part2}
Recombination excitation of the [N~{\sc ii}], [O~{\sc ii}] and [O~{\sc iii}] 
auroral lines}

In NGC\,7009, most N and O atoms are in their doubly ionized stages. 
Recombination of N$^{2+}$ and O$^{2+}$ can be important in exciting the weak 
[N~{\sc ii}] $\lambda$5754 auroral line and the [O~{\sc ii}] auroral and 
nebular lines $\lambda\lambda$7320, 7330, and $\lambda\lambda$3726, 3729, 
leading to apparent high electron temperatures from the ($\lambda$6548 + 
$\lambda$6584)/$\lambda$5754 and ($\lambda$7320 + $\lambda$7330)/$\lambda$3727 
ratios (Rubin \citealt{rubin1986}). Similarly, the [O~{\sc iii}] auroral line 
$\lambda$4363 could be also enhanced by recombination, leading to an 
overestimated $T_\mathrm{e}$. Here we estimate the contribution by 
recombination using the empirical fitting formulae [Equations (1), (2) and (3)
in their paper] given by Liu et al. \cite{liu2000}, which are derived from the
recombination coefficients and transition data for the meta-stable levels of 
N$^+$, O$^+$ and O$^{2+}$ ions. The three equations are as follows:

\begin{equation}
\label{recomb.nii}
\frac{I_\mathrm{R}(\lambda~5754)}{I(\rm{H}\beta)} =
3.19t^{0.30}\times\frac{\rm{N}^{2+}}{\rm{H}^+}, (0.5\leq~t~\leq2.0),
\end{equation}

\begin{equation}
\label{recomb.oii}
\frac{I_\mathrm{R}(\lambda~7320 + \lambda~7330)}{I(\rm{H}\beta)} =
9.36t^{0.44}\times\frac{\rm{O}^{2+}}{\rm{H}^+}, (0.5\leq~t~\leq1.0),
\end{equation}
and

\begin{equation}
\label{recomb.oiii}
\frac{I_\mathrm{R}(\lambda~4363)}{I(\rm{H}\beta)} =
12.4t^{0.59}\times\frac{\rm{O}^{3+}}{\rm{H}^+},
\end{equation}
where $t$ = $T_\mathrm{e}$/10$^4$~K.

For the [N~{\sc ii}] $\lambda$5754 auroral line, we estimate the 
recombination-contributed intensity of about 0.037 [$I(\rm{H}\beta)$ = 100], 
which amounts to 9.5 per cent of the observed intensity of $\lambda$5754. Here 
we use the N$^{2+}$/H$^+$ abundance ratio derived from the [N~{\sc iii}] 
57$\mu$m far-IR line obtained from the ISO/LWS observation by Liu et al. 
\cite{liu2001}. The temperature range that Eq.~(4) can be applied is 
5000~$\leq\,T_\mathrm{e}\,\leq$~20,000~K. Here we adopt the average 
$T_\mathrm{e}$ of NGC\,7009 deduced from CEL diagnostic ratios, which is about 
10,020~K (Table~\ref{diagnostics}), to calculate the contribution by 
recombination. If instead we use the N$^{2+}$/H$^+$ abundance ratio derived 
from the UV N~{\sc iii}] $\lambda$1747 line, with the observed flux from the 
IUE observation by Perinotto \& Benvenuti \cite{pb1981}, then we find that the
recombination contributes about 4.5 per cent to the total intensity of 
$\lambda$5754. We assume that the N$^{2+}$/H$^+$ abundance from far-IR line is
more reliable. Subtracting the flux contributed by the recombination 
excitation, we obtain a ($\lambda$6584 + $\lambda$6548)/$\lambda$5754 ratio 
that is about 9.5 per cent higher than the value before the correction, and 
that brings the resultant electron temperature from original 12,100~K down to 
now 10,780~K, assuming an electron density of 10,000~cm$^{-3}$. If we use the 
average electron density 4290~cm$^{-3}$, the corrected [N~{\sc ii}] 
nebular-to-auroral line ratio then yields a temperature of 11,520~K.

For the [O~{\sc ii}] auroral lines $\lambda\lambda$7320, 7330, we estimate the
recombination excitation contributes an intensity of 0.303 [$I(\rm{H}\beta)$ 
= 100], which is about 13 per cent of their total intensity. Here the 
O$^{2+}$/H$^+$ abundance ratio derived from [O~{\sc iii}] optical CELs is 
used. And we use an electron temperature of 9810~K, derived from the 
[O~{\sc iii}] $\lambda$4959/$\lambda$4363 ratio. Subtracting the recombination 
contribution, we obtain a temperature of 9850~K from the corrected [O~{\sc ii}]
($\lambda$7320 + $\lambda$7330)/$\lambda$3729 ratio, with electron density 
assumed to be 10,000~cm$^{-3}$. The result seems reasonable, given the fact 
that O$^+$ resides in a relatively low ionization region.

For the [O~{\sc iii}] auroral line $\lambda$4363, we estimate the 
recombination excitation contributes an intensity of 0.049 [$I(\rm{H}\beta)$ =
100], which amounts to only about 0.68 per cent of the total $\lambda$4363 
intensity. This contribution is negligible. Here the O$^{3+}$/H$^+$ abundance 
ratio derived from the O~{\sc iv}] $\lambda$1403 UV line is used, with its 
observed flux from the IUE observations by Perinotto \& Benvenuti 
\cite{pb1981}, and electron temperature is set to be 10,020~K.

For the above three cases, the forbidden line temperatures and CEL abundances
are used to calculate the intensities contributed by recombination excitation.
These results are applicable for a chemically uniform nebula, and can be used
to explain the apparently high temperatures yielded by the [N~{\sc ii}] 
$\lambda$5754 and [O~{\sc ii}] $\lambda\lambda$7320, 7330 auroral lines. For a 
chemically inhomogeneous nebula, for example, the model proposed by Liu et al. 
\cite{liu2000} to explain the long-existing discrepancies in the electron 
temperature diagnostics and heavy element abundance determinations in PNe and 
probably also in H~{\sc ii} regions, using recombination lines/continua on the 
hand and CELs on the other, the approach above may not be applicable. In their 
bi-abundance model, Liu et al. \cite{liu2000} suggest that the nebula contains 
a ``cold", H-deficient and metal-rich component, where most of the observed 
fluxes of heavy element ORLs arise. The CELs are mainly emitted by the hot 
ambient plasma. Plasma diagnostics with the aid of newly calculated effective 
recombination coefficients during the past decade shows that this ``cold" 
component has electron temperatures much lower than the forbidden line 
temperatures by nearly an order of magnitude, reaching, in some PNe, as low as
1000~K or even lower (Liu et al. \citealt{liu2006a}). The heavy element 
abundances of this ``cold" component are much higher than those derived from 
CELs by a factor of 2 to 10, and up to 2 orders of magnitude in the most 
extreme case (Liu et al. \citealt{liu2006a}). Evidence in favor of this 
bi-abundance model has been provided by a number of ORL surveys in the past 
decade.

The ``cold" and H-deficient (metal-rich) component postulated by Liu et al. 
\cite{liu2000} may also contribute to the auroral lines of [N~{\sc ii}] and 
[O~{\sc ii}]. If we adopt an electron temperature for the ``cold" component of 
plasma, say 1000~K, and N$^{2+}$/H$^+$, O$^{2+}$/H$^+$ and O$^{3+}$/H$^+$ 
abundance ratios derived from ORLs, then the contribution of recombination 
excitation to the [N~{\sc ii}] $\lambda$5754 line is 0.061 [$I(\rm{H}\beta)$ = 
100], about 15.6 per cent of its observed value, higher than the percentage of
9.5 derived above using the N$^{2+}$/H$+$ abundance ratio derived from the 
[N~{\sc iii}] 57$\mu$m. For the [O~{\sc ii}] $\lambda\lambda$7320, 7330 
auroral lines, the contribution is 22 per cent, also much higher than the value 
estimated above. For the [O~{\sc iii}] $\lambda$4363 auroral line, the 
recombination excitation contributes about 15 per cent to its total intensity, 
if $T_\mathrm{e}$ = 1000~K and the O$^{3+}$/H$^+$ abundance ratio derived 
from the O~{\sc iii} M8 $\lambda$3265 line is adopted.

From the discussion above, it is clear that, depending on the physical models 
assumed, the exact amounts of the recombination contribution to the 
[O~{\sc ii}] and [N~{\sc ii}] auroral lines remain uncertain, and thus the 
electron temperatures derived from them (and consequently ionic abundances 
derived from CELs). The difficulty is that, in the scenario of the bi-abundance
model, without resort detailed modeling, it is impossible to separate the 
contributions from the two components of plasma of vastly different physical 
conditions and chemical composition to the observed fluxes of emission lines, 
CELs or ORLs likewise.

\subsection{\label{plasma_diagnostics:part3}
$T_\mathrm{e}$ and $N_\mathrm{e}$ from the H~{\sc i} Balmer recombination 
spectrum}

Together with the temperatures and densities derived from CEL ratios, 
Table~\ref{diagnostics} also gives the Balmer jump temperature derived from 
the ratio of the nebular continuum Balmer discontinuity at 3646\,{\AA}\, to 
H11 $\lambda$3770, which is defined as 
[$I_\mathrm{c}(\lambda3643)$~$-$~$I_\mathrm{c}(\lambda3681)$]/$I$(H11). Here 
$I_\mathrm{c}(\lambda3643)$ and $I_\mathrm{c}(\lambda3681)$ are the nebular 
continua at 3643 and 3681\,{\AA}, respectively. Using the fitting formula from 
Liu et al. \cite{liu2000},

\begin{equation}
\label{BJTe}
T_\mathrm{e} = 368\times(1 + 0.259~\frac{\rm He^+}{\rm H^+} + 
3.409~\frac{\rm He^{2+}}{\rm H^+})(\frac{\rm BJ}{{\rm H}11})^{-3/2}~{\rm K},
\end{equation}
we derive a Balmer jump temperature of 6490~K (Table~\ref{diagnostics}). 
Here He$^+$/H$^+$ = 0.099 and He$^{2+}$/H$^+$ = 0.013, as derived, 
respectively, from the He~{\sc i} $\lambda\lambda$4471, 5876 and 6678 and from
the He~{\sc ii} recombination line $\lambda$4686. Our Balmer jump temperature 
agrees within the uncertainties with that obtained by Zhang et al. 
\cite{zhang2004}, but is much lower than the values 8000$\sim$8300~K given by 
Liu et al. \cite{liu1995}. The difference could not be caused by the different 
extinction values adopted, given the small wavelength gap between the Balmer 
discontinuity and H~11, but is more likely due to different nebular regions 
being sampled. The Balmer jump temperature 6500~K is about 3300~K lower than 
the [O~{\sc iii}] forbidden line temperature $T_\mathrm{e}$([O~{\sc iii}]), 
and about 3500~K lower than the value of 10,020~K, yielded by a number of 
optical CEL ratios (Table~\ref{diagnostics}).

\begin{figure}
\begin{center}
\epsfig{file=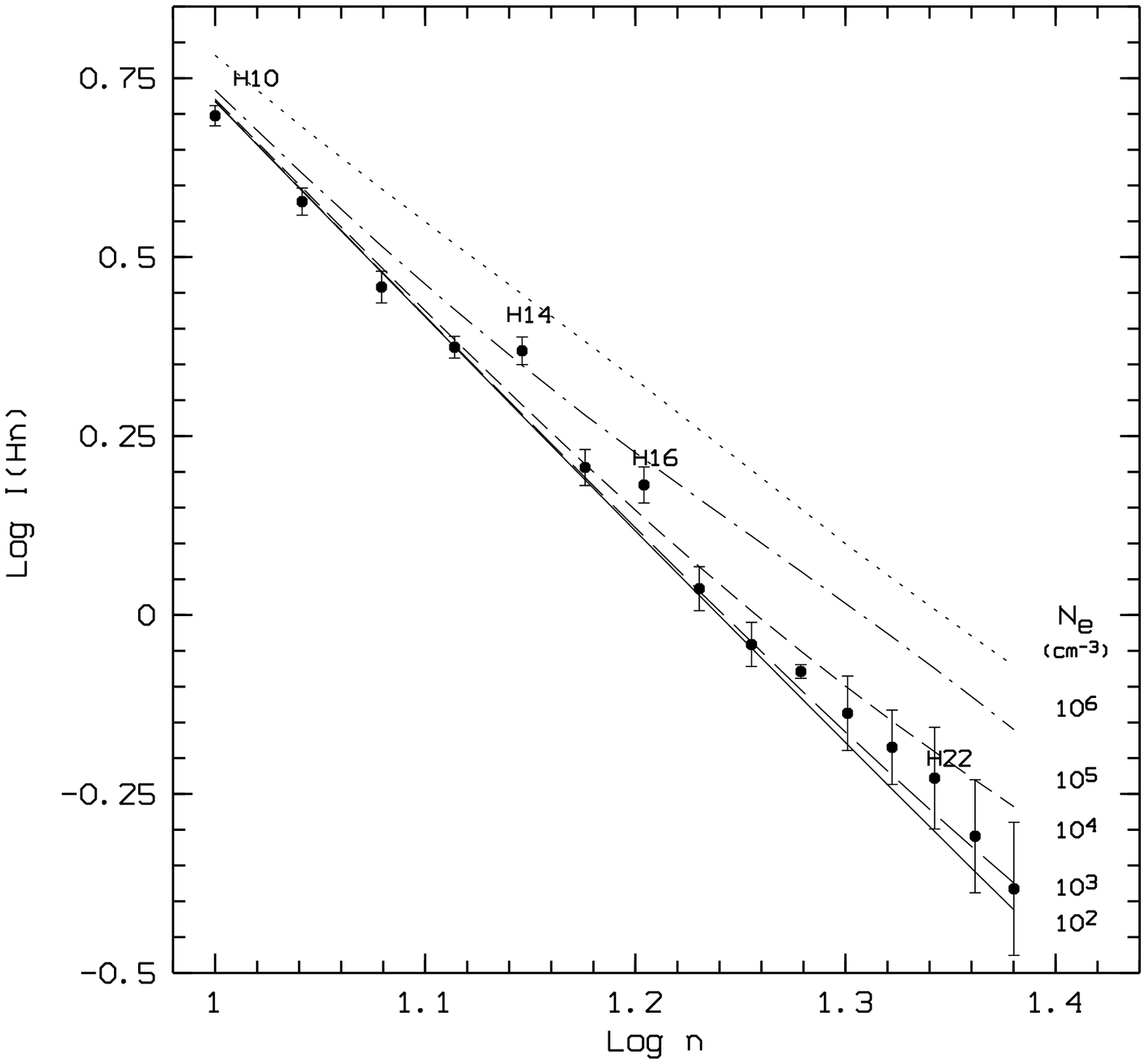,width=7cm,angle=-90}
\caption{Observed intensities (in units where H$\beta$ = 100) of high-order 
Balmer lines H\,$n$($n$ = 10, 11, 12, ..., 24) as a function of the principal 
quantum number $n$. The intensities of H14 and H16 may be overestimated due to
blending of unknown lines, in addition to those deblended (see text for more 
details). The curves show the predicted Balmer decrements for a range of 
electron density from $N_\mathrm{e}$ = 10$^2$ to 10$^6$~cm$^{-3}$. A constant 
temperature of 6500~K, derived from the nebular continuum Balmer discontinuity,
has been assumed in all cases.}
\label{Balmer_decrement}
\end{center}
\end{figure}

The intensities of the high-order Balmer lines relative to H$\beta$, 
$I(n\rightarrow\,2, n\geq\,10)$, are sensitive to electron density 
$N_\mathrm{e}$ and thus provide a valuable density diagnostic. Unlike the 
Balmer discontinuity, this diagnostic is insensitive to the adopted electron 
temperature and can be used to probe the presence of high-density plasmas 
($N_\mathrm{e}\geq\,10^6$~cm$^{-3}$). With our spectral resolution, the Balmer 
decrements can be measured up to $n\sim\,23$. For higher $n$, the fluxes become
unreliable due to line blending. Fig.\,\ref{Balmer_decrement} shows the Balmer
line intensities as a function of $n$ for 10~$\leq\,n\leq$~24. Also overplotted
in the Figure are the predicted Balmer line intensities for different densities
for a fixed $T_\mathrm{e}$ of 6500~K deduced from the Balmer discontinuity. 
The predictions are calculated from the H~{\sc i} emissivities of Storey \& 
Hummer \cite{sh1995}. An optimization yields a Balmer decrement density of 
3000~cm$^{-3}$ (Table~\ref{diagnostics}). This value is about 3300~cm$^{-3}$ 
lower than Zhang et al. \cite{zhang2004}. 

\begin{figure}
\begin{center}
\epsfig{file=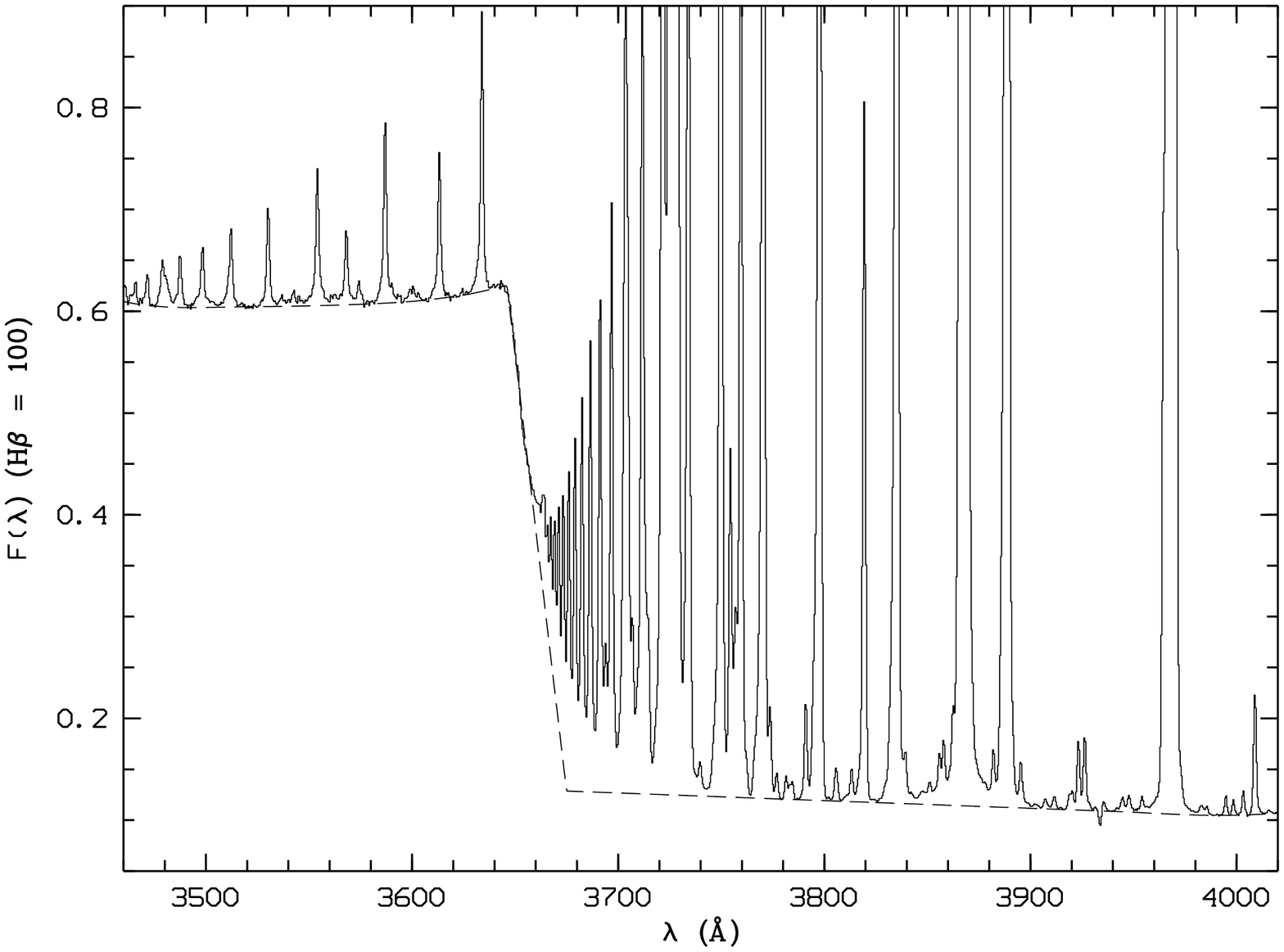,width=6cm,angle=-90}
\caption{The near-UV spectrum of NGC\,7009 from 3460 to 4020\,{\AA}, showing 
the nebular continuum Balmer discontinuity at 3646\,{\AA}. The ratio of the 
Balmer jump to H\,11 $\lambda$3770 yields an electron temperature of 6490~K, 
about 3300~K lower than that derived from the [O~{\sc iii}] 
$\lambda$4959/$\lambda$4363 nebular-to-auroral line ratio. The dashed line is 
a two-part fit to the continuum, bluewards and redwards of the Balmer jump. 
The observed continuum level includes a small contribution from the central 
star. The stellar continuum is expected to be smooth over the plotted 
wavelength range and thus should not affect the magnitude of the observed 
Balmer discontinuity. The spectrum has been corrected to the rest wavelengths 
but not corrected for extinction, and is normalized such that $F({\rm H}\beta)$ 
= 100.}
\label{Balmer_jump}
\end{center}
\end{figure}

Given the crowdedness of lines in the spectral region redwards of the Balmer 
discontinuity, where in addition to the high-order Balmer lines also fall many
faint ORLs from helium and heavy elements, there are no line-free spectral 
windows from the Balmer discontinuity to 3745\,{\AA}, as is shown in 
Fig.\,\ref{Balmer_jump}. The continuum level in this spectral range is 
therefore estimated by linear extrapolation from longer wavelengths 
(Fig.\,\ref{Balmer_jump}), a method adopted by Liu et al. \cite{liu2000} in 
their determination of the Balmer discontinuity of the NGC~6153 spectrum. 
Given the short wavelength range covered (from the leftward of the Balmer 
discontinuity to H~11 $\lambda$3770) and the flatness of the nebular continuum,
the local continuum level thus derived should be secure enough. As estimated 
by Liu et al. \cite{liu2000} for NGC~6153, the resultant Balmer line 
intensities could be accurate to a few per cent.

Multi-Gaussian fitting is performed in the continuum-subtracted spectrum to 
derive individual line fluxes, especially for the wavelength region near the 
Balmer discontinuity, where many ORLs from He~{\sc i}, He~{\sc ii} and heavy 
element ions are blended with the crowd of high order H~{\sc i} Balmer lines. 
H\,14 $\lambda$3721.94 is partially blended with the [O~{\sc ii}] $\lambda$3726,
in addition to the [S~{\sc iii}] 3p$^2$~$^1$S$_{0}$ -- 3p$^2$~$^3$P$_{1}$ 
$\lambda$3721.69 line, the latter contributes about 30 per cent to the total 
intensity of H\,14 as estimated from the multi-Gaussian fitting. A high-order 
He~{\sc ii} line 28g~$^2$G -- 4f~$^2$F$^{\rm o}$ $\lambda$3720.41 probably 
also contributes to H\,11, at the level less than 1 per cent. H\,15 
$\lambda$3711.97 is blended with two Ne~{\sc ii} lines M1 $\lambda$3709.62 and
M5 $\lambda$3713.08, and two O~{\sc iii} M14 lines $\lambda$3714.03 and 
$\lambda$3715.08 also affect the wing of H\,15. H\,16 $\lambda$3703.85 is 
partially blended with the O~{\sc iii} $\lambda$3707.25, in addition to the 
O~{\sc iii} M14 $\lambda$3702.75 and He~{\sc i} M25 7d~$^3$D -- 
2p~$^3$P$^{\rm o}$ $\lambda$3705.00. H\,17 $\lambda$3697.15 and H\,18 
$\lambda$3691.55 are both affected by the Ne~{\sc ii} M1 
3p~$^4$P$^{\rm o}_{5/2}$ -- 3s~$^4$P$_{5/2}$ $\lambda$3694.21. For low-order 
Balmer lines H\,$n$ ($n$ = 10, 11, 12), the deblending of faint ORLs is 
relatively easy as the spectrum becomes less crowded. 
Fig.\,\ref{Balmer_decrement} shows that H\,14 and H\,16 may be overestimated 
whereas those of H\,23 and H\,24 underestimated.

The electron density derived from the high-order Balmer lines is 
3000~cm$^{-3}$, which does not differ much from the average density 
4290~cm$^{-3}$ derived from the optical CEL ratios of [O~{\sc ii}], 
[S~{\sc ii}], [Cl~{\sc iii}] and [Ar~{\sc iv}] (Table~\ref{diagnostics}).

\subsection{\label{plasma_diagnostics:part4}
$T_\mathrm{e}$ and $N_\mathrm{e}$ from the H~{\sc i} Paschen recombination 
spectrum}

Electron temperature is also estimated from the Paschen decrements, using

\begin{figure}
\begin{center}
\epsfig{file=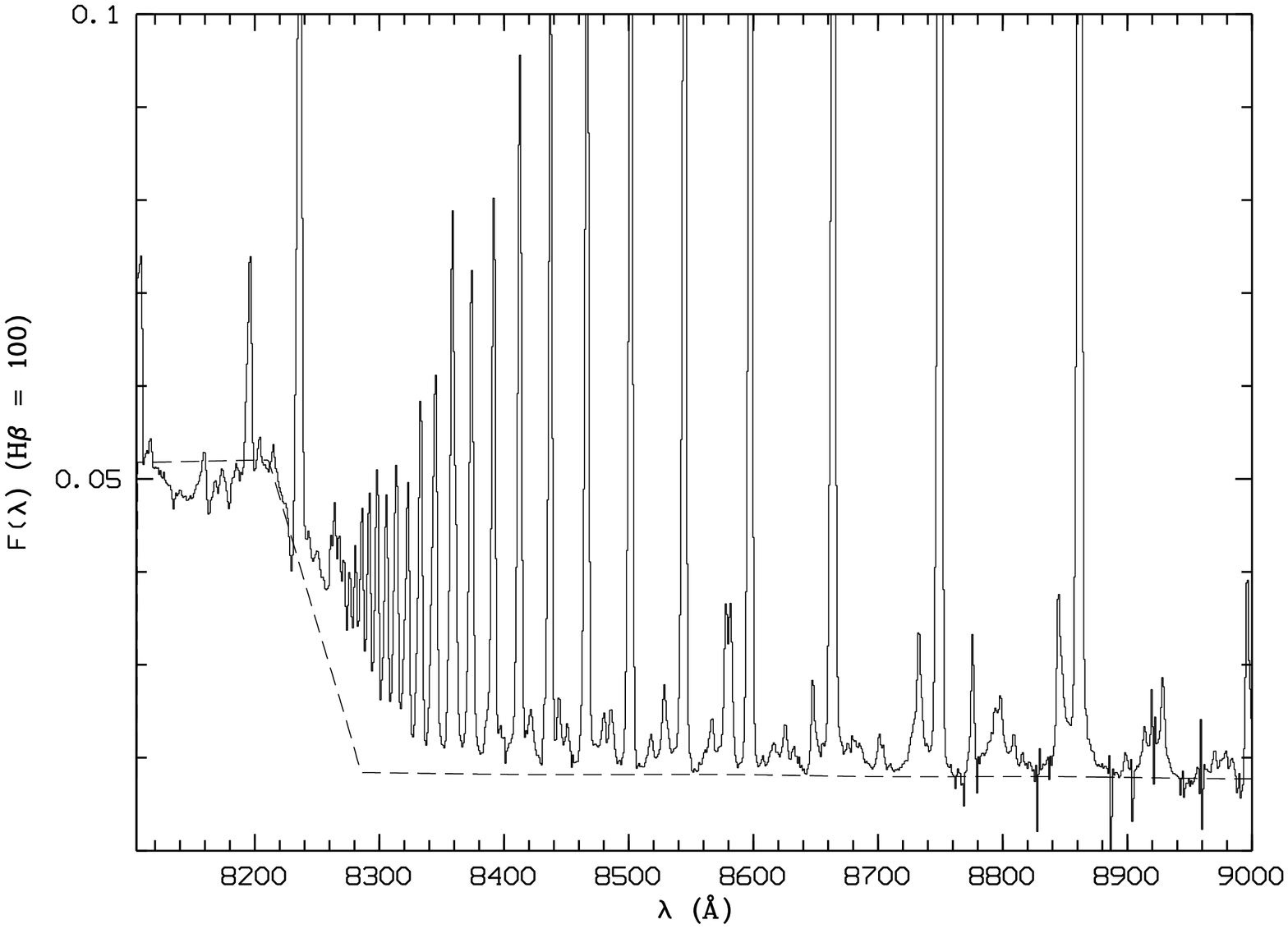,width=6cm,angle=-90}
\caption{The red spectrum of NGC\,7009 from 8105 to 9000\,{\AA}, showing the 
nebular continuum Paschen discontinuity at 8204\,{\AA}, as well as the Paschen
decrements. The dashed line is a two-part fit to the continuum, bluewards and 
redwards of the Paschen jump. The observed continuum level includes a small 
contribution from the central star. The stellar continuum is expected to be 
extremely smooth over the plotted wavelength range and thus does not affect 
the estimate of the Paschen discontinuity. The spectrum has been corrected to 
the rest wavelengths but not corrected for extinction, and is normalized such 
that $F({\rm H}\beta)$ = 100.}
\label{Paschen_jump}
\end{center}
\end{figure}

\begin{figure*}
\begin{center}
\epsfig{file=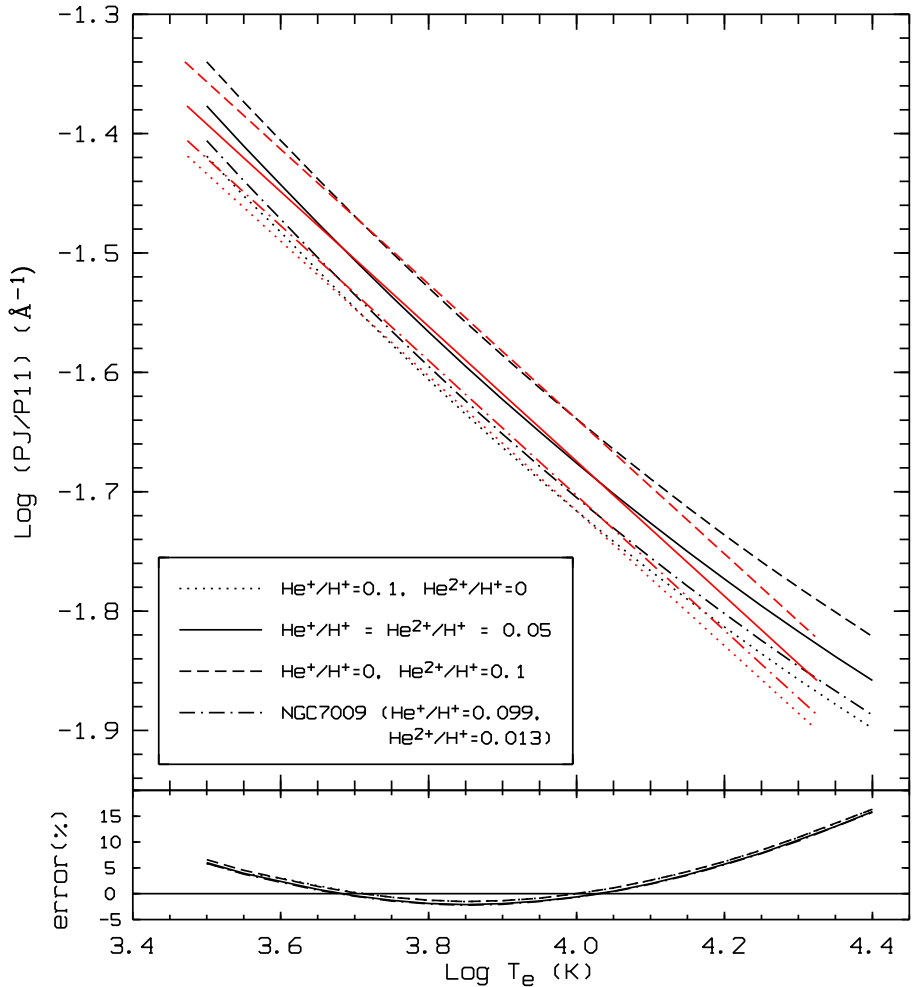,width=12cm,angle=0}
\caption{$Upper~Panel$: The Paschen discontinuity to P11 ratio, 
PJ/P11~$\equiv$~[$I_{\rm c}(8194)$ $-$ $I_{\rm c}(8269)$]/$I$(P11), as a 
function of $T_\mathrm{e}$, in logarithm space, for (a) He$^+$/H$^+$ = 0.1 and 
He$^{2+}$/H$^+$ = 0 (dotted line), (b) He$^+$/H$^+$ = He$^{2+}$/H$^+$ = 0.05 
(solid line), (c) He$^+$/H$^+$ = 0 and He$^{2+}$/H$^+$ = 0.1 (dashed line), 
and (d) He$^+$/H$^+$ = 0.099 and He$^{2+}$/H$^+$ = 0.013 (dash-dot line), 
which is the case of NGC\,7009. $I_{\rm c}(8194)$ and $I_{\rm c}(8269)$ are the 
nebular continuum fluxes at 8194 and 8269\,{\AA}, respectively, which bracket 
the H~{\sc i} Paschen discontinuity at 8204\,{\AA}\, and several 
discontinuities of He~{\sc i} and He~{\sc ii}. Also overplotted are the red 
curves calculated from Eq.\,\ref{Te_PJ} for the four cases. $Lower~Panel$: 
Distribution of fitting errors using Eq.\,\ref{Te_PJ} as a function of 
$T_\mathrm{e}$ for the four cases as defined in the $upper~panel$. Here the 
fitting error is calculated from the real unit of electron temperature (K), 
not in logarithm space as in the $upper~panel$.}
\label{PJ.vs.Te}
\end{center}
\end{figure*}

\begin{equation}
\label{Te_PJ}
T_{\rm e} = 8.72\times(1 + 0.52\frac{{\rm He}^{+}}{{\rm H}^{+}} + 
   4.40\frac{{\rm He}^{2+}}{{\rm H}^{+}})(\frac{\rm PJ}{\rm P~11})^{-1.77}.
\end{equation}
where PJ/P11 is ($I_{\rm c}(8194)$~$-$~$I_{\rm c}(8169)$)/$I$(P11), in units 
of {\AA}$^{-1}$. The fitting errors given by Eq.\,\ref{Te_PJ} are less than 5
per cent for the temperature range
10$^{3.55}$\,$\leq$\,Log$T_\mathrm{e}$\,$\leq$\,10$^{4.15}$, which is from 
3550 to 14,000~K, and are less than 16 per cent for 
10$^{3.50}$\,$\leq$\,Log$T_\mathrm{e}$\,$\leq$\,10$^{4.40}$, as is shown in 
Fig.\,\ref{PJ.vs.Te}. The electron temperature derived from the Paschen jump
for NGC\,7009 is 6750$\pm$160~K (Table~\ref{diagnostics}).

\begin{figure}
\begin{center}
\epsfig{file=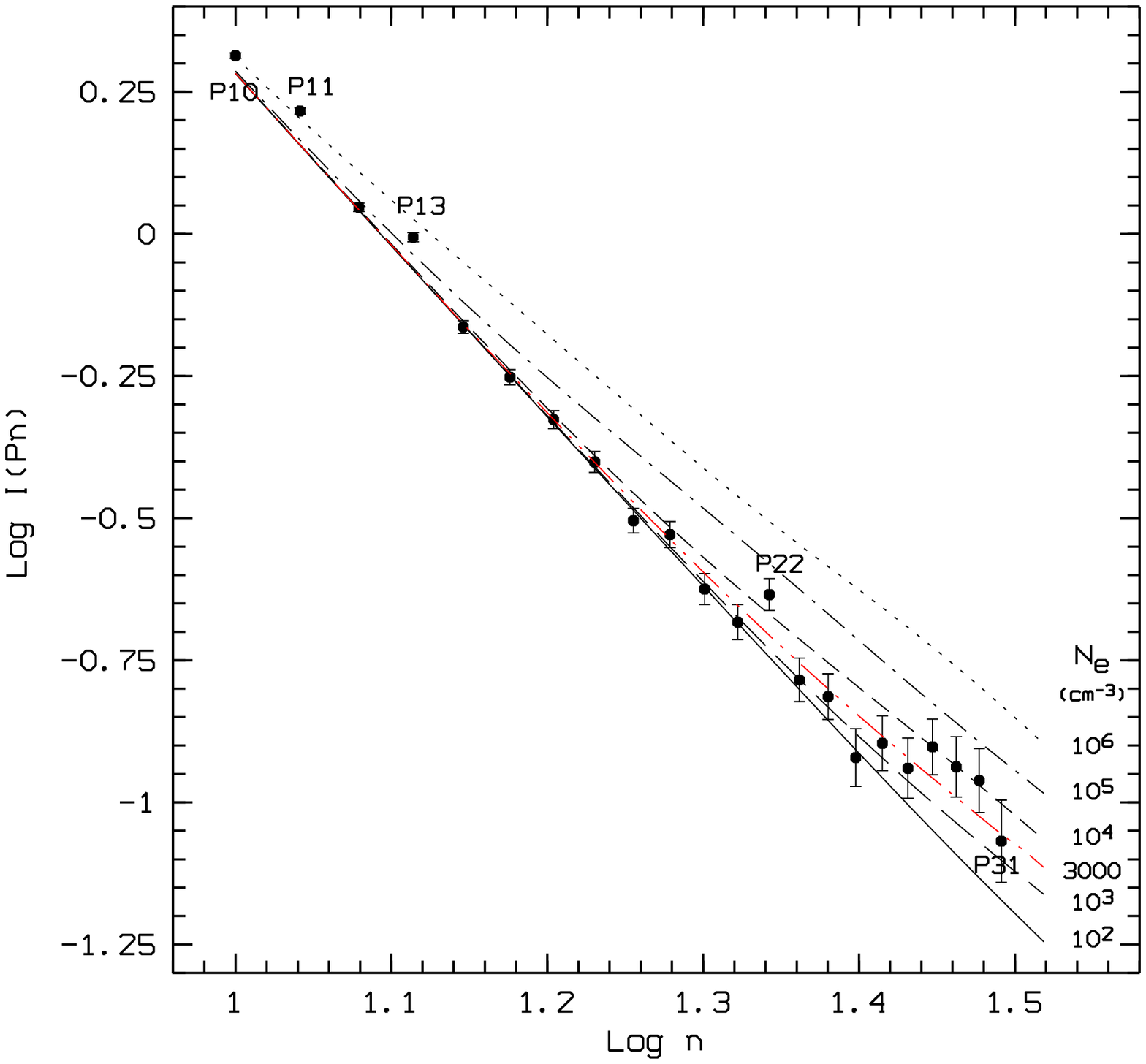,width=7cm,angle=-90}
\caption{Observed intensities (in units where H$\beta$ = 100) of high-order 
Paschen lines P\,$n$ ($n$ = 10, 11, 12,..., 31) as a function of the principal 
quantum number $n$. The intensities of P11 and P13 are overestimated due to 
the blend of other emission lines. Also probably overestimated are Paschen 
lines with $n\geq\,28$. The various curves show respectively the predicted 
Paschen decrements for electron densities from 10$^2$ to 10$^6$~cm$^{-3}$. A 
constant temperature of 6750~K as derived from the nebular continuum Paschen 
discontinuity has been assumed in all cases.}
\label{Paschen_decrement}
\end{center}
\end{figure}

\subsection{\label{plasma_diagnostics:part5}
$T_\mathrm{e}$'s from the He~{\sc i} and He~{\sc ii} recombination spectra}

%
Electron temperatures are derived from the intensity ratios of the He~{\sc i} 
$\lambda$4471, $\lambda$5876, $\lambda$6678 and $\lambda$7281 recombination 
lines, as is shown in Table~\ref{diagnostics}. Zhang et al. \cite{zhang2005b} 
developed an electron temperature diagnostic with the He~{\sc i} recombination 
lines, utilizing the He~{\sc i} line emissivities calculated by Benjamin et 
al. \cite{benjamin1999}. Analytic formulae for the He~{\sc i} line 
emissivities were used:

\begin{equation}
\label{hei_fit}
\frac{I_{1}}{I_{2}} = 
\frac{a_{1}}{a_{2}} t^{b_{1} - b_{2}} {\rm exp}(\frac{c_{1} - c_{2}}{t}),
\end{equation}
where $t$ = $T_\mathrm{e}$/10$^4$~K, and $a_{i}$, $b_{i}$ and $c_{i}$ are the 
fitting parameters for emissivities given by Benjamin et al. 
\cite{benjamin1999}. Eq.\,(\ref{hei_fit}) is valid for the temperature 
5000$\sim$20,000~K. For lower temperatures ($T_\mathrm{e}\,<\,5000$~K), Zhang
et al. \cite{zhang2005b} obtained similar fits using the He~{\sc i} 
recombination line intensities calculated by Smits \cite{smits1996} and the 
effects of collisional excitation from the 2s $^3$S and 2s $^1$S meta-stable 
levels by Sawey \& Berrington \cite{sb1993}.

Using the fit formula~(\ref{hei_fit}), we derived an electron temperature of 
5100$\pm$400~K from the He~{\sc i} $\lambda$7281/$\lambda$6678 line ratio. We 
also derived 3850$\pm$350~K from the $\lambda$7281/$\lambda$5876 line ratio, 
and 4360$\pm$2200~K from the $\lambda$5876/$\lambda$4471 line ratio. Here the 
uncertainties were estimated based on the line ratio errors which are deduced 
from the line fitting errors. The large uncertainty of 
$T_\mathrm{e}$($\lambda$5876/$\lambda$4471) is mainly due to the complex line 
blending of $\lambda$4471 (as mentioned in the next paragraph). The electron 
temperatures derived from the $\lambda$7281/$\lambda$6678 and 
$\lambda$7281/$\lambda$5876 ratios agree well with those given by Zhang et al.
\cite{zhang2005b} within errors. The He~{\sc i} $\lambda$6678/$\lambda$4471 
line ratio gives an electron temperature of about 9700~K, which is higher than
the values given by the other three He~{\sc i} line ratios by nearly a factor 
of 2.

The high electron temperature yielded by the He~{\sc i} 
$\lambda$6678/$\lambda$4471 ratio is probably questionable. The He~{\sc i} 
$\lambda$4471 line suffers the most from line blending among the four 
He~{\sc i} lines: it is blended with two O~{\sc ii} M86c lines $\lambda$4469.46
and $\lambda$4469.48, three O~{\sc iii} lines M49c $\lambda$4471.02, 
M49c $\lambda$4475.17 and M45b $\lambda$4476.11, and one Ne~{\sc ii} line M61b
$\lambda$4468.91. In addition, both wings of the $\lambda$4471 line are 
affected by weak features. All these complication bring notable uncertainties 
to the intensity of the $\lambda$4471 line. The $\lambda$4471 line intensity 
adopted in the current analysis has been corrected for the contributions from 
the blended lines listed above, using the effective recombination coefficients 
available for the O~{\sc ii} and O~{\sc iii} lines. We estimate that all those 
blended lines may contribute at least 10 per cent of the observed intensity of 
the $\lambda$4471 line. If we do not consider the effects of those blended 
O~{\sc ii}, O~{\sc iii} and Ne~{\sc ii} lines, and assume the measured 
intensity of the feature is entirely from the $\lambda$4471 line, the 
$\lambda$6678/$\lambda$4471 ratio would then yield an electron temperature of 
17,500~K. We conclude that the electron temperature given by the 
$\lambda$6678/$\lambda$4471 line ratio is probably unreliable.

It is generally assumed that the He~{\sc i} Lyman lines (1s$n$p~$^1$P$^{\rm o}$
-- 1s$^2$~$^1$S) are optically thick and that the He~{\sc i} singlet 
transitions follow Case~B recombination theory under nebular conditions. 
Under this assumption, all the Lyman photons emitted by transitions from upper
levels with $n\,\ge\,3$ of singlet He~{\sc i} will be re-absorbed and scattered
by He$^{0}$, and eventually converted to He~{\sc i} Ly\,$\alpha$ at 584\,{\AA}
plus photons at longer wavelengths, in particular those of the series 
$n$p~$^1$P$^{\rm o}$ -- 2s~$^1$S. Of the three singlet series represented by 
$\lambda$7281 (3s~$^1$S -- 2p~$^1$P$^{\rm o}$), $\lambda$5016 
(3p~$^1$P$^{\rm o}$ -- 2s~$^1$S) and $\lambda$6678 (3d~$^1$D -- 
2p~$^1$P$^{\rm o}$) lines, the $n$d~$^1$D -- 2p~$^1$P$^{\rm o}$ lines are 
essentially unaffected by optical depth effects as well as the assumption of 
Case~A or Case~B recombination, while the other two series, 
$n$p~$^1$P$^{\rm o}$ -- 2s~$^1$S and $n$s~$^1$S -- 2p~$^1$P$^{\rm o}$, are 
strongly affected (Liu et al. \citealt{liu2001}). The predicted Case~A 
intensities of the $n$p~$^1$P$^{\rm o}$ -- 2s~$^1$S lines, relative to 
He~{\sc i} 2p~$^1$P$^{\rm o}$ -- 4d~$^1$D $\lambda$4922, are more than an 
order of magnitude lower than those predicted in Case~B. For the $n$s~$^1$S 
-- 2p~$^1$P$^{\rm o}$ series, the predicted Case~A intensities are nearly a 
factor of 2 lower than for Case~B. In the current measurements, the 
$\lambda$7281/$\lambda$4922 intensity ratio is 0.389, which is quite close 
to the Case~B value 0.391 of Brocklehurst \cite{brocklehurst1972}. The 
intensity of 4s~$^1$S -- 2p~$^1$P$^{\rm o}$ $\lambda$5048, relative to the 
$\lambda$4922 line, also agrees with the Case~B prediction within the errors.
Our measured intensity ratio of He~{\sc i} 3p~$^1$P$^{\rm o}$ -- 2s$^1$S 
$\lambda$5016 to the $\lambda$4922 line is 1.324, which is about 30 per cent 
lower than the Case~B value 1.860 of Brocklehurst \cite{brocklehurst1972}, 
and is higher than the Case~A prediction by a factor of 30. The observed 
$\lambda$6678/$\lambda$4922 line ratio agrees with the predictions of 
Brocklehurst \cite{brocklehurst1972} within the errors. Thus the singlet 
transitions of He~{\sc i} in NGC\,7009 are probably close to the Case~B 
assumption. Here an electron temperature of 5000~K and a density of 
10$^4$~cm$^{-3}$ are assumed for the theoretical predictions.

Liu et al. \cite{liu2001} raised the possibility of departure from Case~B of 
the He~{\sc i} singlet lines: It is possible that the He~{\sc i} Lyman photons
(with $\lambda\leq$\,584\,{\AA}) are destroyed by photoionization of neutral
hydrogen H$^{0}$ and or absorption by dust grains, rather than re-absorbed by 
He$^{0}$ and converted to photons at longer wavelengths, if there is a 
significant amount of neutral hydrogen or dust grains co-existing with 
He$^{+}$. These processes effectively cause a departure of the He~{\sc i} 
singlet recombination spectrum from Case~B towards Case~A, even though the 
nebula is optically thick to the He~{\sc i} Lyman lines. Thus the electron 
temperature derived from the $\lambda$7281/$\lambda$6678 and 
$\lambda$7281/$\lambda$5876 lines could be underestimated, due to a decrease of
the $\lambda$7281 intensity. As analyzed above, this departure is not clearly 
observed in our spectrum. However, detailed photoionization modeling is 
needed to test the hypothesis and deduce the contribution of this mechanism 
quantitatively.

Given that the $\lambda$7281/$\lambda$6678 ratio is probably the best 
He~{\sc i} line ratio suitable for electron temperature determinations 
(Zhang et al. \citealt{zhang2005b}), we adopt 5100~K deduced from the ratio 
as the He~{\sc i} emission line temperature.

\begin{figure}
\begin{center}
\epsfig{file=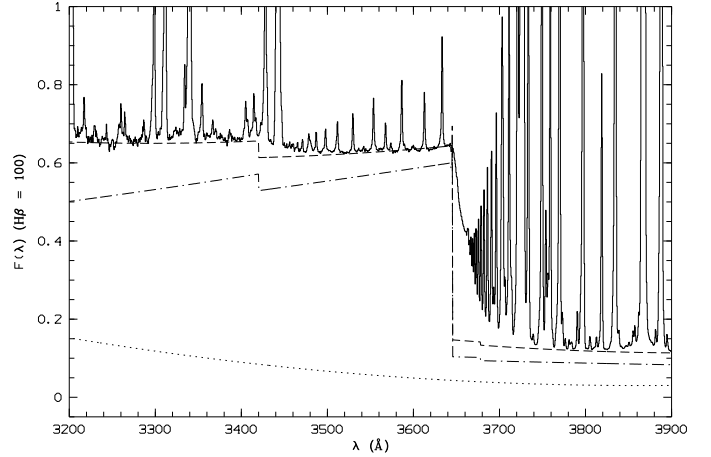,width=6cm,angle=-90}
\caption{CCD spectrum (solid line) of NGC\,7009 from 3200 to 3900\,{\AA}, 
showing the He~{\sc i} discontinuity at 3421\,{\AA}, as well as the Balmer 
discontinuity at 3646\,{\AA}. The intensity is normalized such that H$\beta$ =
100, and corrected for extinction. The dashed line is the total continuum, the
sum of the theoretical nebular continuum (dot-dashed line) and the scattered 
stellar light (dotted line).}
\label{HeI_3421jump}
\end{center}
\end{figure}

%
The He~{\sc i} discontinuity at 3421\,{\AA}, which is formed by singly ionized 
helium recombining to the 2~$^3$P$^{\rm o}$ term of He~{\sc i}, is detected 
(Fig.\,\ref{HeI_3421jump}), thanks to the high S/N of the spectrum. Theoretical
calculations are carried out to fit the spectrum near the He~{\sc i} and 
H~{\sc i} Balmer discontinuities. The detailed description of the fitting 
procedure is in Zhang et al. \cite{zhang2004}, and the same method has been 
used by Zhang et al. \cite{zhang2009} to simultaneously determine electron 
temperatures of PNe from the H~{\sc i} and He~{\sc i} discontinuities. The 
current fits yield an electron temperature of 7800$\pm$200~K 
(Table~\ref{diagnostics}), which is lower than those given by Liu et al. 
\cite{liu1993}: 11,720$^{+2590}_{-2260}$~K (PA = 360$^{\rm o}$) and 
10,400$^{+2480}_{-2440}$~K (PA = 270$^{\rm o}$). The error of the temperature 
is estimated from repeated runs of the fitting procedure. The actual 
uncertainty in temperature could be even larger.

\begin{figure*}
\begin{center}
\epsfig{file=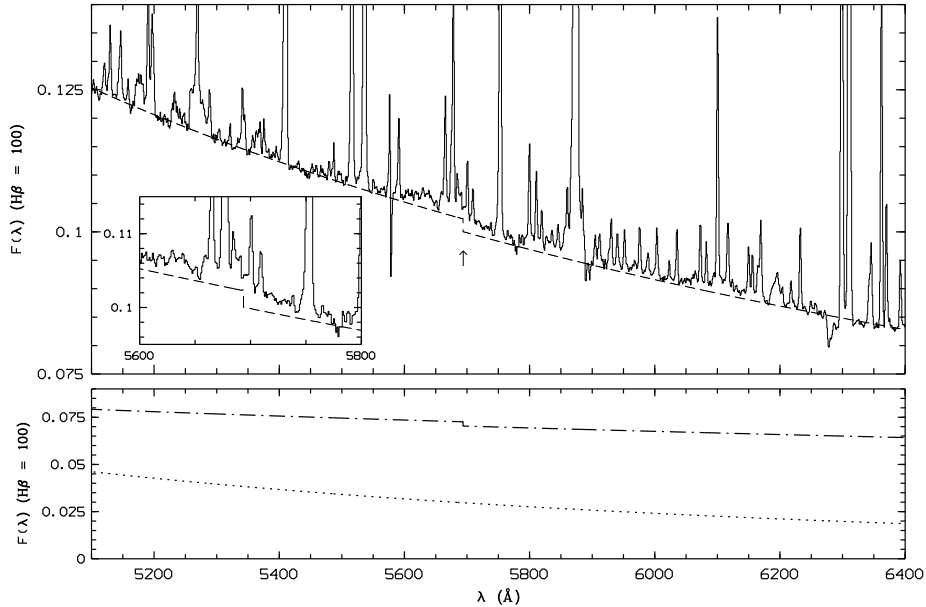,width=8cm,angle=-90}
\caption{$Upper~Panel$: CCD spectrum of NGC\,7009 (solid line) from 5100 to 
6400\,{\AA}, showing the He~{\sc ii} discontinuity at 5694\,{\AA}\, marked by 
an arrow. The insert is a zoom of the jump. The dashed line is a fit to the 
continuum level of the spectrum. $Lower~Panel$: The theoretical nebular 
continuum (dot-dashed line) and scattered stellar light (dotted line). The 
lower panel is not on the same scale as the upper one.}
\label{HeII_5694jump}
\end{center}
\end{figure*}

The He~{\sc ii} jump at 5694\,{\AA}, which is formed by doubly ionized helium 
recombining to spectral terms of He$^{+}$ with principal quantum number $n$ = 
5, is also detected (Fig.\,\ref{HeII_5694jump}), although it is not obvious. 
Spectral fitting to the He~{\sc ii} discontinuity at 5694\,{\AA} gives an 
electron temperature of about 11,000~K, with an uncertainty of about 2000~K. 
This estimate is slightly lower than 13,800$^{+6200}_{-3800}$~K given by Liu 
et al. \cite{liu1993}. This fitting procedure is carried out in a broad 
wavelength range, 5100 -- 6400\,{\AA}\, (Fig.\,\ref{HeII_5694jump}), due to the 
weakness of the discontinuity.

\section{\label{discussion}
Discussion}

\subsection{\label{discussion:part1}
Flux errors}

The spectrum of NGC\,7009 presented in the current work is among the deepest 
CCD spectra ever taken for an emission line nebula. The high S/N's coupled with
the medium resolution make it difficult to find a line-free region to estimate
the local continuum and its uncertainties, and consequently the uncertainties 
of the fluxes of individual emission lines. Here we have adopted the flux 
errors yielded by the multi-Gaussian fitting, and they are listed in the last 
column of Table~\ref{linelist}.

The errors vary but only have a weak dependence on wavelength. They do have a 
strong dependence on the strength of the line, as expected. Table~\ref{errors}
shows the average flux errors for different flux bins. Also presented in 
Table~\ref{errors} are the numbers of the emission lines in each flux bins. 
For lines with fluxes higher than $\log[F(\lambda)/F({\rm H}\beta)]>-5.0$, 
those with flux errors larger than 100$\%$ are excluded in the calculation of 
the average error of individual flux bin, and their numbers are given in the 
notes to Table~\ref{errors}. 

Amongst the large number of lines detected or deblended in the spectrum of 
NGC\,7009, of particular interests are ORLs from C~{\sc ii}, N~{\sc ii}, 
O~{\sc ii} and Ne~{\sc ii}, and they will be analyzed separately in a 
subsequent paper. Those ORLs have typical fluxes about 10$^{-4}$ to 10$^{-2}$ 
of H$\beta$, with typical measurement uncertainties of 10 to 20 per cent 
(Table~\ref{errors}). The best observed ORLs of N~{\sc ii} and O~{\sc ii} have
errors that are well below 10 per cent: e.g., the fitted intensity of 
O~{\sc ii} M1 3p~$^4$D$^{\rm o}_{7/2}$ -- 3s~$^4$P$_{5/2}$ $\lambda$4649.13, 
which is shown in Fig.\,\ref{4625-4680}, is 0.666 (on a scale where H$\beta$ 
is 100), with a fitting error of less than 2 per cent; the fitted intensity of 
N~{\sc ii} M3 3p~$^3$D$_{3}$ -- 3s~$^3$P$^{\rm o}_{2}$ $\lambda$5679.56 is 
0.136 (H$\beta$ = 100), and its fitting error is about 5 per cent. The 
Ne~{\sc ii} M2 3p~$^4$D$^{\rm o}_{7/2}$ -- 3s~$^4$P$_{5/2}$ $\lambda$3334.84 
line has a measured intensity of 0.428 (H$\beta$ = 100), with a fitting error
of 8 per cent. Accurate measurements of those ORLs are of paramount importance
for plasma diagnostics and abundance determinations using ORLs.

\begin{table}
\centering
\caption{Flux measurement errors.}
\label{errors}
\begin{tabular}{crcc}
\hline
Flux range & Number   & The average & Note\\
           & of lines & error (\%)  &     \\
\hline
$-$5.0 $<$ $\log[F(\lambda)/F({\rm H}\beta)]$ $<$ $-$4.0 & 312 & $>$28.0 & (1)\\
$-$4.0 $<$ $\log[F(\lambda)/F({\rm H}\beta)]$ $<$ $-$3.0 & 627 & 19.3    & (2)\\
$-$3.0 $<$ $\log[F(\lambda)/F({\rm H}\beta)]$ $<$ $-$2.0 & 157 & 10.6    & (3)\\
$-$2.0 $<$ $\log[F(\lambda)/F({\rm H}\beta)]$ $<$ $-$1.0 & 64  &  6.34   & (4)\\
$-$1.0 $<$ $\log[F(\lambda)/F({\rm H}\beta)]$ $<$    0   & 15  &  1.56   &    \\
     0 $<$ $\log[F(\lambda)/F({\rm H}\beta)]$ $<$    2.0 & 5   & $<$0.1  &    \\
\hline
\end{tabular}
\begin{description}
\item [(1)] 93 lines with errors larger than 100\%.
\item [(2)] 28 lines with errors larger than 100\%.
\item [(3)] 2 lines with errors larger than 100\%.
\item [(4)] 2 lines with errors larger than 100\%.
\end{description}
\end{table}

\subsection{\label{discussion:part2}
Number of lines}

\subsubsection{\label{discussion:part2:a}
Cumulative numbers}

\begin{figure*}
\begin{center}
\epsfig{file=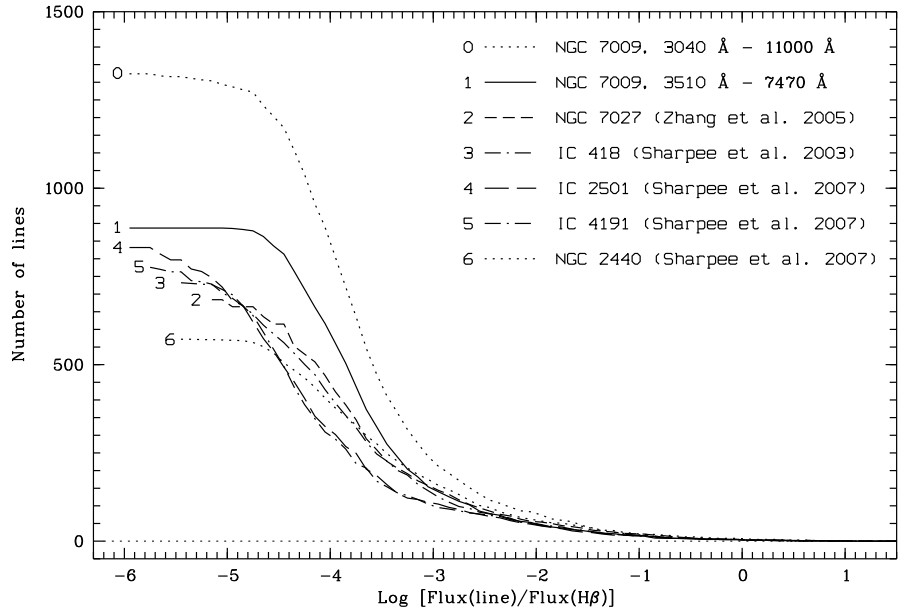,width=8cm,angle=-90}
\caption{Cumulative number of emission lines exceeding a given flux level 
identified in the spectrum of NGC\,7009 and in several recently published 
deep spectra of other PNe. For the curves labeled with numbers 1, 2, 3, 4, 5 
and 6, we consider only lines within the wavelength range 3510 -- 7470\,{\AA},
which is the common range covered by the observations of these PNe. For curve 
1, the dubious identifications in NGC\,7009 are not included. The dotted 
curve 0 includes all emission lines of NGC\,7009 in the whole wavelength 
range from 3040 to 11,000\,{\AA}, except those 235 alternative identifications
given by {\sc emili}.}
\label{cumulative}
\end{center}
\end{figure*}

The current spectrum is among the deepest ever taken for a PN, which allows 
the detection or deblending of many weak ORLs that are valuable for nebular 
analyses. In this subsection, a statistical analysis of the distribution of 
the emission lines as a function of intensity is presented. 
Fig.\,\ref{cumulative} shows the cumulative numbers of lines exceeding a given
flux level relative to H$\beta$. For comparison, also shown in the Figure are 
numbers of emission lines detected in deep spectroscopy of a number of PNe 
published in the recent literature: NGC\,7027 (short-dashed curve, labeled 
with number 2; Zhang et al. \citealt{zhang2005a}), IC\,418 (dot-dashed curve, 
labeled with number 3; Sharpee et al. \citealt{sharpee2003}), IC\,2501 
(long-dashed curve, labeled with number 4; Sharpee et al. 
\citealt{sharpee2007}), IC\,4191 (dot-dot-dashed curve, labeled with number 5;
Sharpee et al. \citealt{sharpee2007}), NGC\,2440 (dotted curve, labeled with 
number 6; Sharpee et al. \citealt{sharpee2007}). In Fig.\,\ref{cumulative}, 
for curves labeled by the numbers from 1 to 6, we consider only those lines 
within the wavelength range 3510 -- 7470\,{\AA} which is covered by all the 
spectra. The curve of cumulative number of lines measured in NGC\,7009 over 
the whole wavelength range 3040 -- 11,000\,{\AA} is also presented (the dotted
curve, which is labeled by 0).

In the wavelength range 3510 -- 7470\,{\AA}, for the very weak intensity bin 
($-5.0\leq\log[F(\lambda)/F({\rm H}\beta)]\leq-4.0$), 270 lines are measured 
and identified in NGC\,7009, 251 in NGC\,7027 (Zhang et al. 
\citealt{zhang2005a}), 418 in IC\,418 (Sharpee et al. \citealt{sharpee2003}) 
and 421, 299 and 166 respectively in the remaining three PNe, IC\,2501, 
IC\,4191 and NGC\,2440 (Sharpee et al. \citealt{sharpee2007}). In the same 
wavelength range, for the medium weak intensity bin 
($-4.0\leq\log[F(\lambda)/F({\rm H}\beta)]\leq-3.0$), NGC\,7009 yields 459 
lines identified, NGC\,7027 311, IC\,418 204 and the other three PNe, 
IC\,2501, IC\,4191 and NGC\,2440 215, 286 and 230, respectively. In the weak 
to strong intensity bin ($\log[F(\lambda)/F({\rm H}\beta)]>-3.0$), NGC\,7009 
gives 153 lines, NGC\,7027 158, IC\,418 105 and the other three PNe, IC\,2501,
IC\,4191 and NGC\,2440 110, 143 and 175, respectively.

In Fig.\,\ref{cumulative}, all curves start to flatten out at 
$\log[F(\lambda)/F({\rm H}\beta)]\sim-4.6$, where the line detection 
incompleteness kicks in, even though lines continue to be detected at fainter 
intensities. In the case of NGC\,7009, no lines are detected with reliable 
identifications with logarithmic intensities lower than $-5.0$. The detection 
limit of the current spectrum of NGC\,7009 is similar to that of NGC\,7027 
(Zhang et al. \citealt{zhang2005a}) but is obviously lower than the deepest 
echelle spectrum ever taken for a PN, i.e. that of IC\,2501 (Sharpee et al. 
\citealt{sharpee2007}).

In the wavelength range 3510 -- 7470\,{\AA}, the total number of emission 
lines identified in NGC\,7009 is 887, larger than in all other PNe shown in 
Fig.\,\ref{cumulative}, including NGC\,7027 (684 lines), NGC\,2440 (572 
lines), IC\,418 (732 lines), IC\,4191 (778 lines) and IC\,2501 (833 lines). 
That is because many optical recombination lines are taken into account and 
are deblended from the spectra of NGC\,7009, although our resolution is much 
lower than those echelle spectra. The total number of the emission lines of 
NGC\,7009, in the complete wavelength range 3040 -- 11,000\,{\AA}, is more 
than 1300, as is shown in Fig.\,\ref{cumulative}. The 235 alternative 
identifications given by {\sc emili} are excluded from the calculation of the 
cumulative line numbers.

\subsubsection{\label{discussion:part2:b}
Emission lines from different ionic species}

In Table~\ref{orl_numbers} we summarize for each ionic species the number of 
permitted lines identified in NGC\,7009. In total 986 permitted lines are 
identified, and most of them are excited mainly by recombination. Similar 
results for forbidden lines (CELs) are presented in Table~\ref{cel_numbers}. 
These include a total of 234 lines. Dubious identifications are excluded from
both tables.

Nearly 200 O~{\sc ii} permitted lines are identified in NGC\,7009. In 
addition, more than 100 N~{\sc ii} and Ne~{\sc ii} lines are identified. The 
identifications of a few Ne~{\sc ii} lines, those from very high-$n$, with 
$n\,\geq\,7$, may be problematic. Further observations are needed to establish
their legitimacy. Only 34 C~{\sc ii} permitted lines are identified. The 
relatively small number of C~{\sc ii} lines is due to the relatively simple 
atomic structure, i.e., an ion with only one valence electron.

Several permitted lines emitted by highly ionized species of the most abundant
heavy elements (C and O) are also detected or deblended in the spectrum of
NGC\,7009: For example, C~{\sc iv} M8 6h~$^2$H$^{\rm o}$ -- 5g~$^2$G 
$\lambda$4658 (please see Fig.\,\ref{4625-4680}) and O~{\sc iv} M2 3d~$^2$D 
-- 3p~$^2$P$^{\rm o}$ $\lambda$3409. Here the C~{\sc iv} $\lambda$4658 line 
may contain a small contamination from the [Fe~{\sc iii}] $\lambda$4658 line 
(Liu et al. \citealt{liu1995}).

\begin{table}
\centering
\caption{Numbers of permitted lines identified for all ionic species. Lines 
with dubious identifications are not included.}
\label{orl_numbers}
\begin{tabular}{lrc}
\hline
Ion & No. of & Note\\
    & lines  & \\
\hline
H~{\sc i}     &  60 & \\
He~{\sc i}    &  93 & \\
He~{\sc ii}   &  65 & \\
C~{\sc i}     &  13 & \\
C~{\sc ii}    &  34 & \\
C~{\sc iii}   &  25 & \\
C~{\sc iv}    &   3 & \\
N~{\sc i}     &  11 & \\
N~{\sc ii}    & 117 & \\
N~{\sc iii}   &  36 & \\
O~{\sc i}     &  17 & \\
O~{\sc ii}    & 192 & \\
O~{\sc iii}   &  47 & \\
O~{\sc iv}    &   3 & \\
Ne~{\sc i}    &  10 & \\
Ne~{\sc ii}   & 135 & \\
Ne~{\sc iii}  &   1 & \\
Na~{\sc i}    &   3 & \\
Mg~{\sc i}    &   5 & \\
Mg~{\sc i}$]$ &   3 & \\
Mg~{\sc ii}   &   2 & \\
Al~{\sc ii}   &   1 & \\
Si~{\sc i}    &   3 & \\
Si~{\sc ii}   &  13 & \\
Si~{\sc iii}  &  12 & \\
Si~{\sc iv}   &   5 & \\
S~{\sc ii}    &   2 & \\
Ar~{\sc i}    &   3 & \\
Ar~{\sc ii}   &   5 & \\
Ca~{\sc i}    &   2 & \\
Fe~{\sc i}    &   4 & \\
Fe~{\sc ii}   &  48 & \\
Fe~{\sc iii}  &  11 & \\
\hline
\end{tabular}
\end{table}

\begin{table}
\centering
\caption{Same as Table~\ref{orl_numbers}, but for CELs.}
\label{cel_numbers}
\begin{tabular}{lrc}
\hline
Ion & No. of & Note\\
    & lines  & \\
\hline
$[$C~{\sc i}$]$    &  2 & \\
$[$N~{\sc i}$]$    &  2 & \\
$[$N~{\sc ii}$]$   &  4 & \\
$[$O~{\sc i}$]$    &  3 & \\
$[$O~{\sc ii}$]$   &  4 & \\
$[$O~{\sc iii}$]$  &  4 & \\
$[$F~{\sc ii}$]$   &  1 & \\
$[$F~{\sc iv}$]$   &  1 & \\
$[$Ne~{\sc iii}$]$ &  4 & \\
$[$Ne~{\sc iv}$]$  &  4 & \\
$[$P~{\sc ii}$]$   &  1 & (1)\\
$[$S~{\sc ii}$]$   &  4 & \\
$[$S~{\sc iii}$]$  &  4 & \\
$[$Cl~{\sc ii}$]$  &  3 & \\
$[$Cl~{\sc iii}$]$ &  4 & \\
$[$Cl~{\sc iv}$]$  &  3 & \\
$[$Ar~{\sc iii}$]$ &  4 & \\
$[$Ar~{\sc iv}$]$  &  4 & \\
$[$Ar~{\sc v}$]$   &  2 & \\
$[$K~{\sc iv}$]$   &  2 & \\
$[$K~{\sc v}$]$    &  1 & \\
$[$K~{\sc vi}$]$   &  1 & \\
$[$Ca~{\sc v}$]$   &  1 & \\
$[$V~{\sc ii}$]$   &  6 & \\
$[$Cr~{\sc ii}$]$  &  6 & \\
$[$Cr~{\sc iii}$]$ &  6 & \\
$[$Cr~{\sc iv}$]$  &  4 & \\
$[$Cr~{\sc v}$]$   &  3 & \\
$[$Mn~{\sc ii}$]$  &  1 & \\
$[$Mn~{\sc iii}$]$ &  1 & \\
$[$Mn~{\sc v}$]$   & 14 & \\
$[$Mn~{\sc vi}$]$  &  2 & \\
$[$Fe~{\sc ii}$]$  & 56 & \\
$[$Fe~{\sc iii}$]$ & 25 & \\
$[$Fe~{\sc iv}$]$  & 15 & \\
$[$Fe~{\sc v}$]$   &  8 & \\
$[$Fe~{\sc vi}$]$  &  8 & \\
$[$Fe~{\sc vii}$]$ &  1 & \\
$[$Co~{\sc ii}$]$  &  1 & \\
$[$Co~{\sc iii}$]$ &  2 & \\
$[$Co~{\sc iv}$]$  &  1 & \\
$[$Co~{\sc v}$]$   &  1 & \\
$[$Ni~{\sc ii}$]$  &  5 & \\
$[$Ni~{\sc iii}$]$ &  2 & \\
$[$Ni~{\sc iv}$]$  &  3 & \\
\hline
\end{tabular}
\begin{description}
\item [(1)] Blended with O~{\sc ii} M89b 4f~D[2]$^{\rm o}_{5/2}$ -- 
3d~$^2$D$_{3/2}$ $\lambda$4669.27, which contributes about 40 per cent to the 
blend at $\lambda$4669.
\end{description}
\end{table}

\begin{figure*}
\begin{center}
\epsfig{file=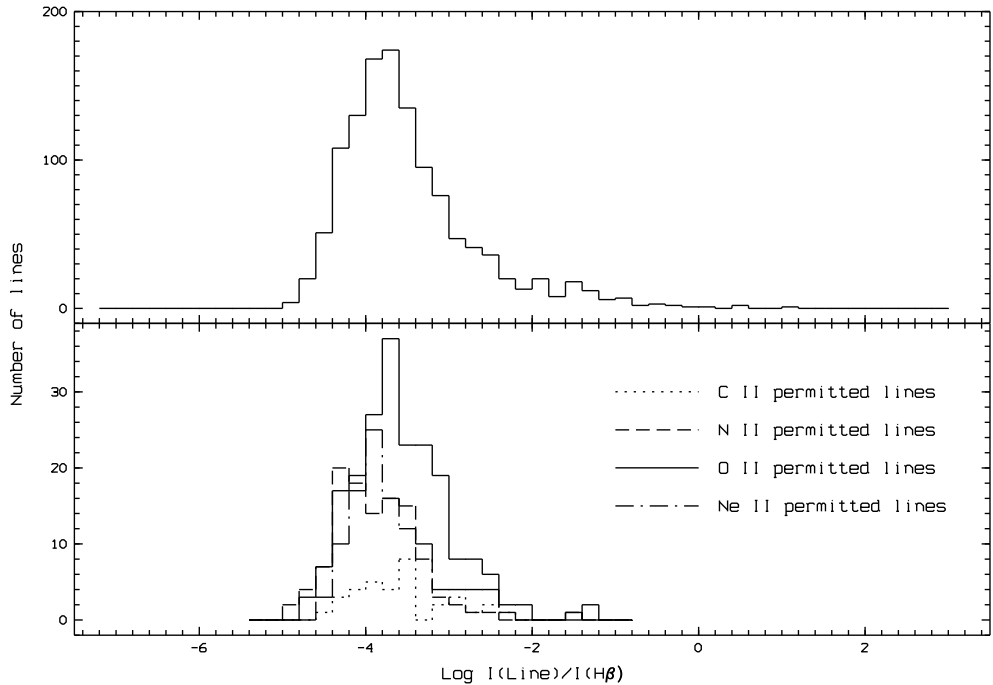,width=9cm,angle=-90}
\caption{$Upper~panel$: Histograms showing the intensity distributions of all 
lines identified. $Lower~panel$: Histograms showing the distributions of 
C~{\sc ii}, N~{\sc ii}, O~{\sc ii} and Ne~{\sc ii} permitted lines. Lines with
dubious identifications are not counted.}
\label{histograms}
\end{center}
\end{figure*}

\begin{figure}
\begin{center}
\epsfig{file=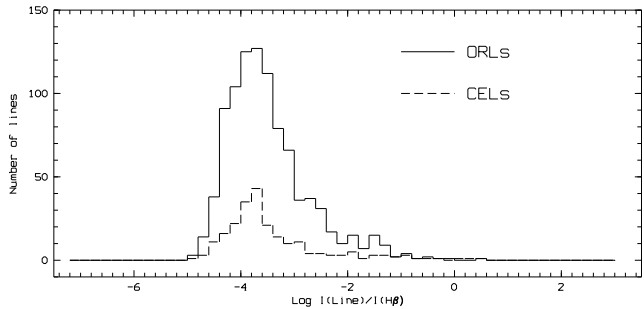,width=4cm,angle=-90}
\caption{Histograms showing the intensity distributions of ORLs and CELs in 
NGC\,7009. Lines with dubious identifications are not counted.}
\label{histo_cel_orl}
\end{center}
\end{figure}

Fig.\,\ref{histograms} shows the histograms of intensity distributions of 
all the lines identified, as well as those of permitted lines from the most 
prominent heavy element ions, C~{\sc ii}, N~{\sc ii}, O~{\sc ii} and 
Ne~{\sc ii}. Fig.\,\ref{histo_cel_orl} shows the distributions of the 
strengths of all identified ORLs and CELs. Most ORLs as well as CELs have 
intensities between 10$^{-5}$ and 10$^{-2}$ of H$\beta$.

\section{\label{summary}
Summary}

Based on very deep, medium-resolution long-slit spectra of the Saturn Nebula 
NGC\,7009, we have detected, deblended and estimated more than 1400 emission 
lines, with the weakest lines estimated down to intensity levels from 
10$^{-6}$ to 10$^{-5}$ that of H$\beta$. In total 1170 emission lines are 
identified, both manually and systematically by the computer-aided code 
{\sc emili}. Multi-Gaussian line profile fitting is carried out across the 
whole wavelength range (3040 -- 11,000\,{\AA}) to retrieve intensities of 
blended features. We first manually identify all the obvious emission features
in the spectra, following the traditional method. Then we use multi-Gaussian 
fitting to obtain fluxes of weak lines that are blended with strong features. 
The most updated atomic transition data are utilized to estimate the fluxes of
those weak lines whose reliable measurements are difficult. Finally, all the 
aforementioned features, except those that are simply estimated from atomic 
data, are further identified with {\sc emili}. All the emission lines 
are presented in a comprehensive table with all detailed transition 
information included.

In addition to singly and doubly ionized species, recombination lines emitted 
by highly ionized ions of C, N and O are also identified. We also detected 
forbidden lines from neutral species C~{\sc i}, N~{\sc i} and O~{\sc i}. The 
majority of lines are in the flux range 10$^{-5}$ -- 10$^{-2}$ that of 
H$\beta$.

Compared to several other PNe that have been extensively studied in the recent
literature, NGC\,7009 exhibits much more faint emission lines, thanks to its 
very rich and prominent ORL spectra from abundant second-row elements. The 
current set of spectra of NGC\,7009 are among the deepest ever taken for a PN.
The detection limit of the current data set is limited by the relatively low 
spectral resolution. Spectroscopy of this bright PN with better resolution 
will no doubt lead to direct detection of more and fainter lines.

\clearpage

\begin{figure*}
\begin{center}
 \epsfig{file=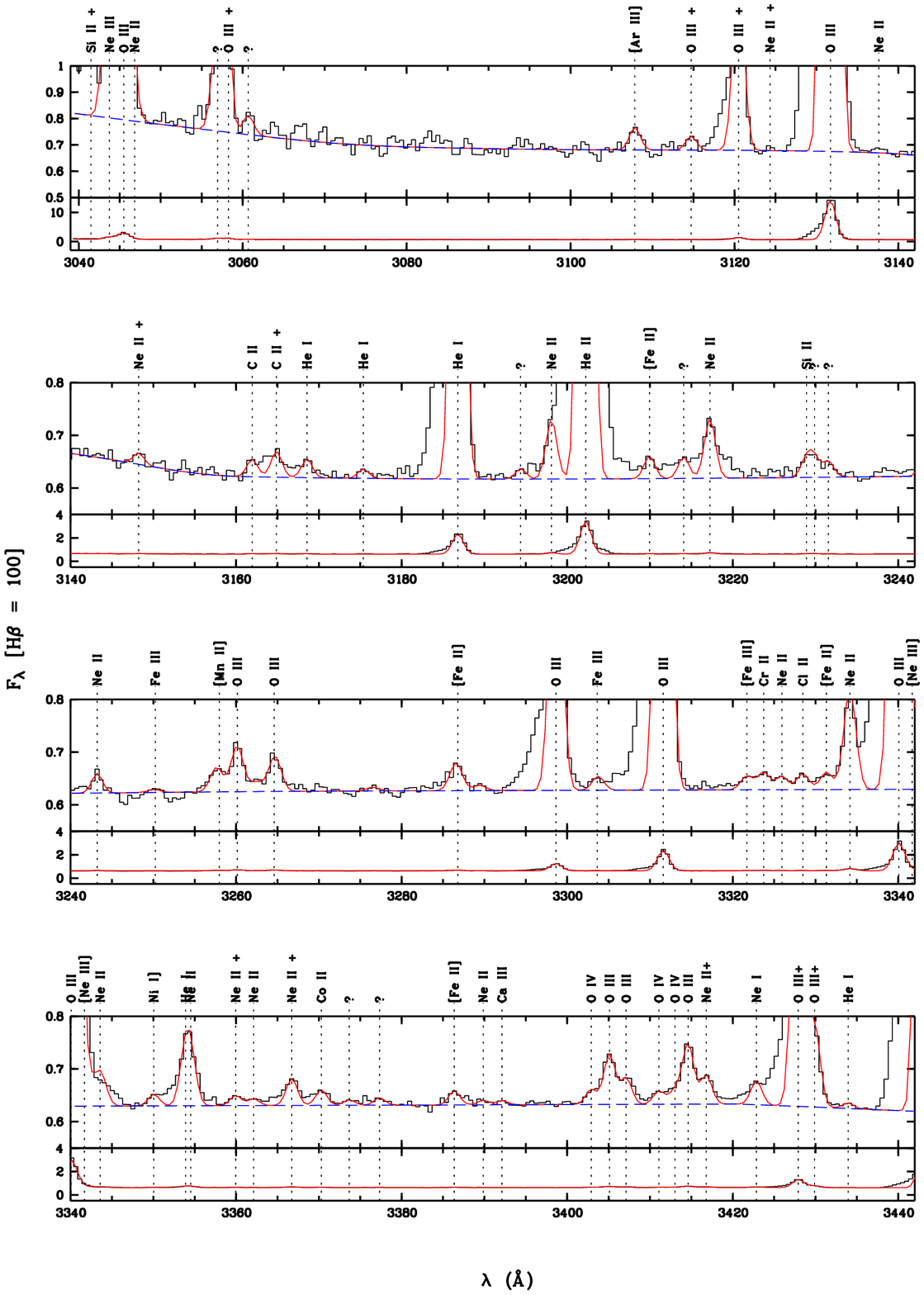,width=15cm,angle=0}
 \caption{Spectrum of NGC\,7009.  A ``$+$" attached to a line identification 
  means that the line is blended with other emission features. The red solid 
  curve is the sum of multi-Gaussian fits to the emission lines; the blue 
  dashed line represents the continuum.}
 \label{spectra_plot}
\end{center}
\end{figure*}

\addtocounter{figure}{-1}
\begin{figure*}
\begin{center}
 \epsfig{file=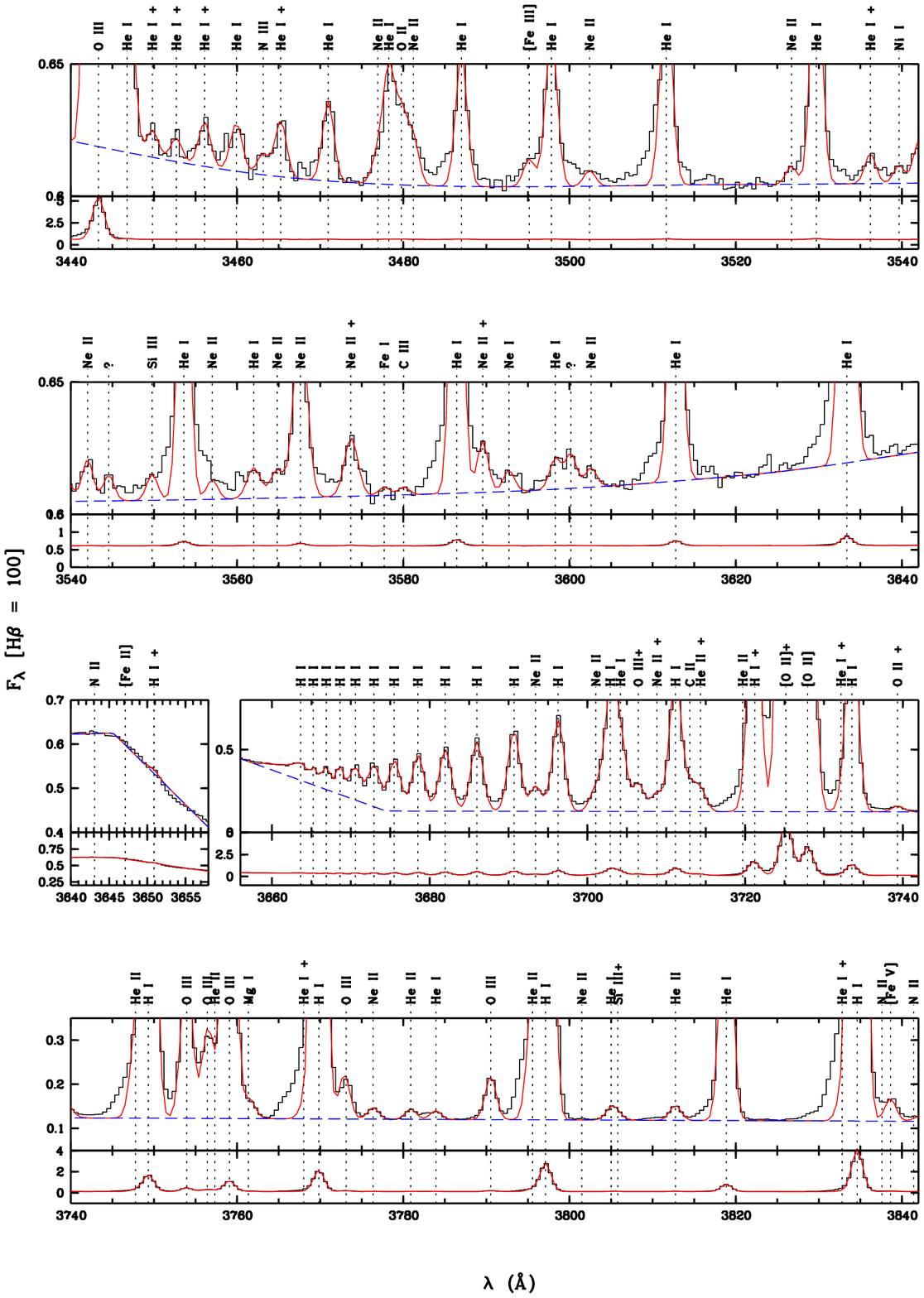,width=15cm,angle=0}
 \caption{Continued.}
 \label{spectra_plot}
\end{center}
\end{figure*}

\addtocounter{figure}{-1}
\begin{figure*}
\begin{center}
 \epsfig{file=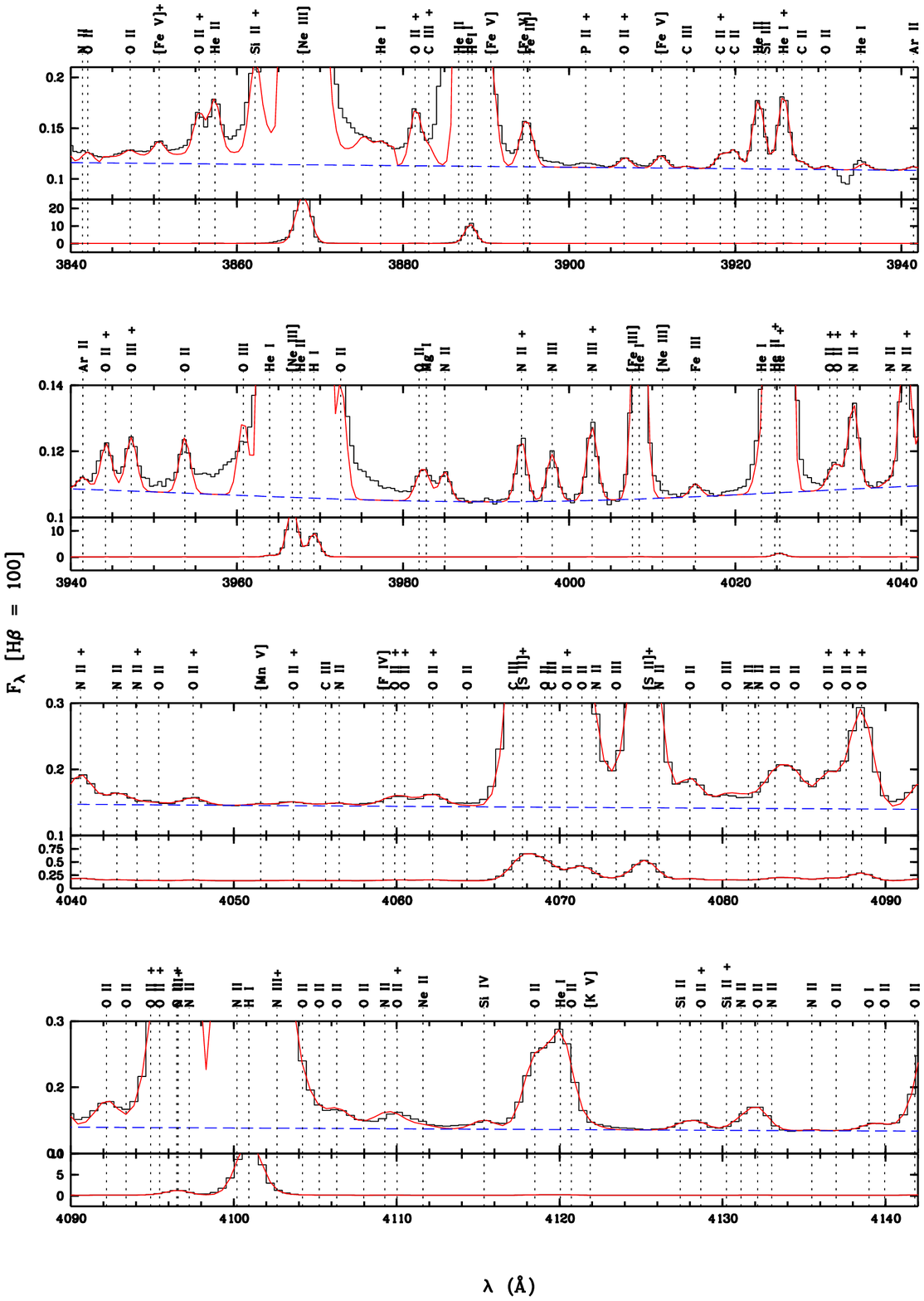,width=15cm,angle=0}
 \caption{Continued.}
 \label{spectra_plot}
\end{center}
\end{figure*}

\addtocounter{figure}{-1}
\begin{figure*}
\begin{center}
 \epsfig{file=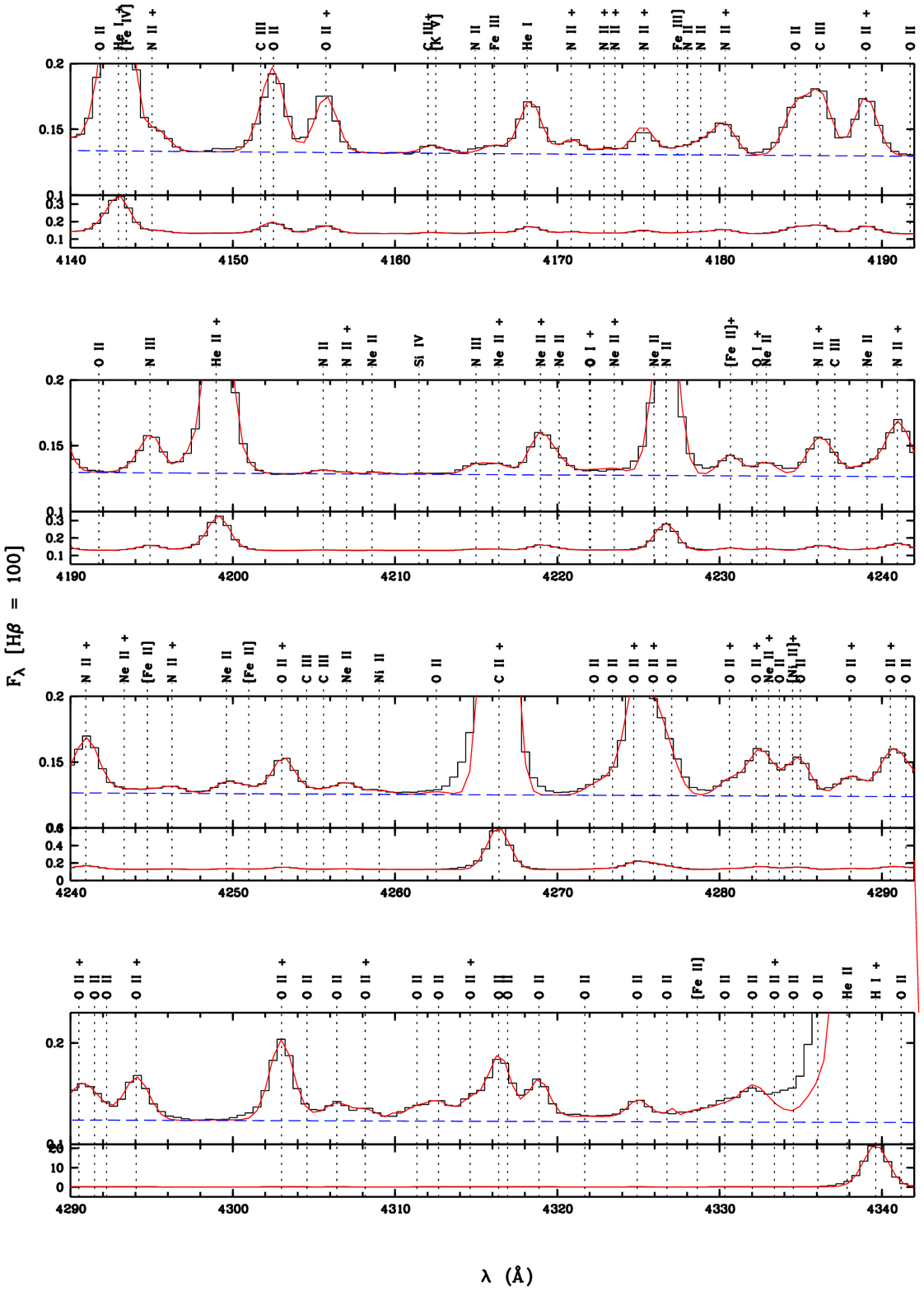,width=15cm,angle=0}
 \caption{Continued.}
 \label{spectra_plot}
\end{center}
\end{figure*}

\addtocounter{figure}{-1}
\begin{figure*}
\begin{center}
 \epsfig{file=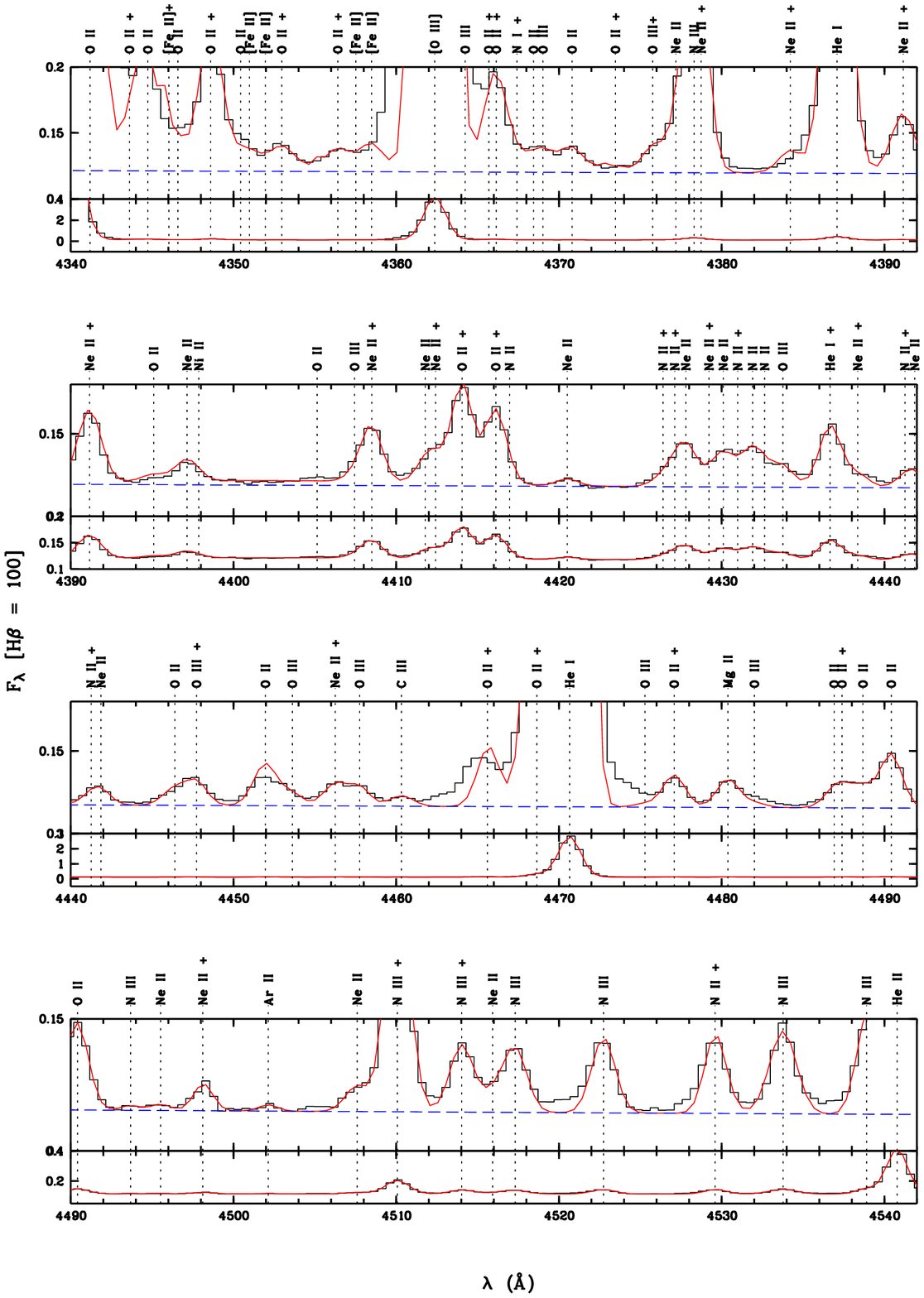,width=15cm,angle=0}
 \caption{Continued.}
 \label{spectra_plot}
\end{center}
\end{figure*}

\addtocounter{figure}{-1}
\begin{figure*}
\begin{center}
 \epsfig{file=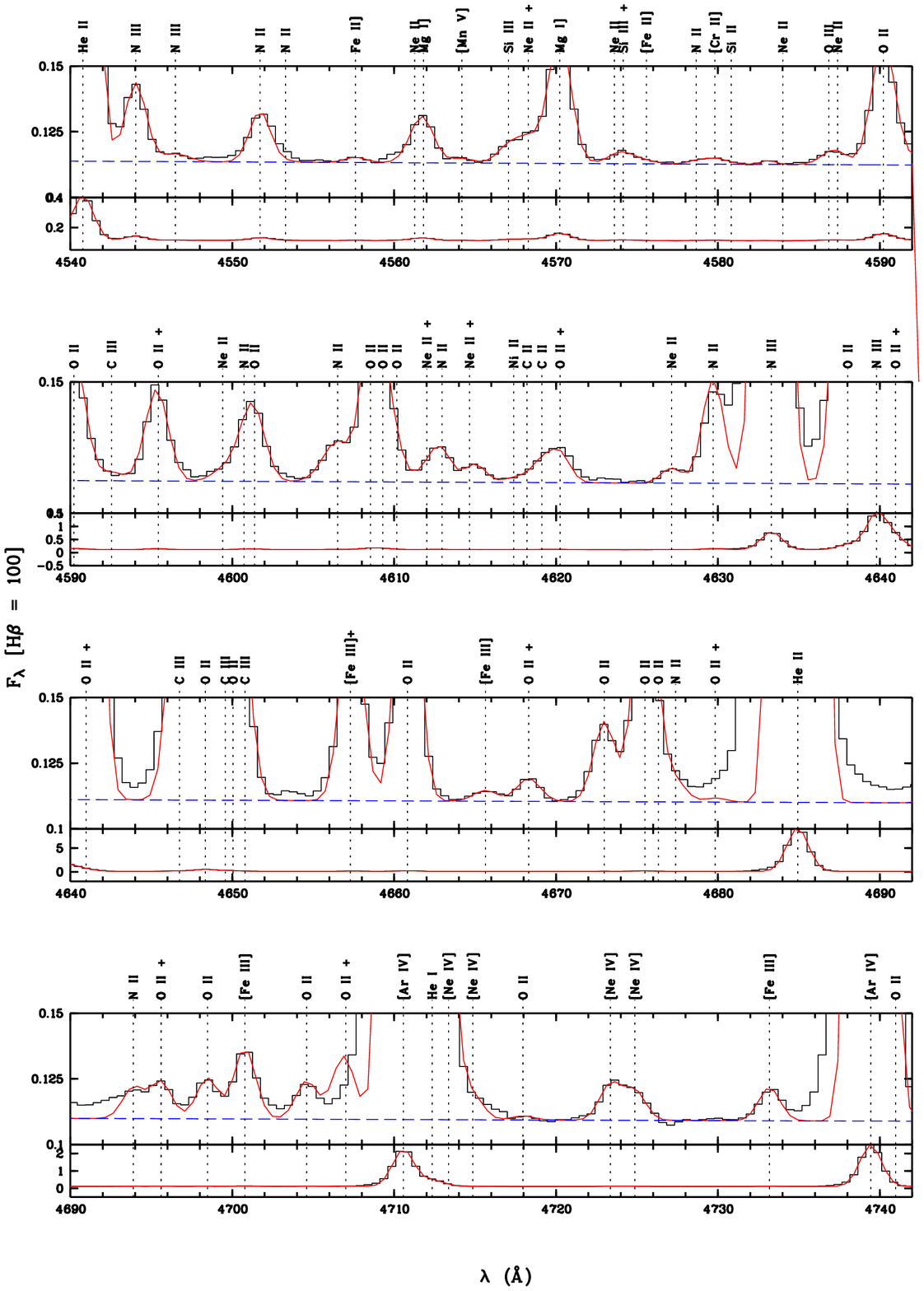,width=15cm,angle=0}
 \caption{Continued.}
 \label{spectra_plot}
\end{center}
\end{figure*}

\addtocounter{figure}{-1}
\begin{figure*}
\begin{center}
 \epsfig{file=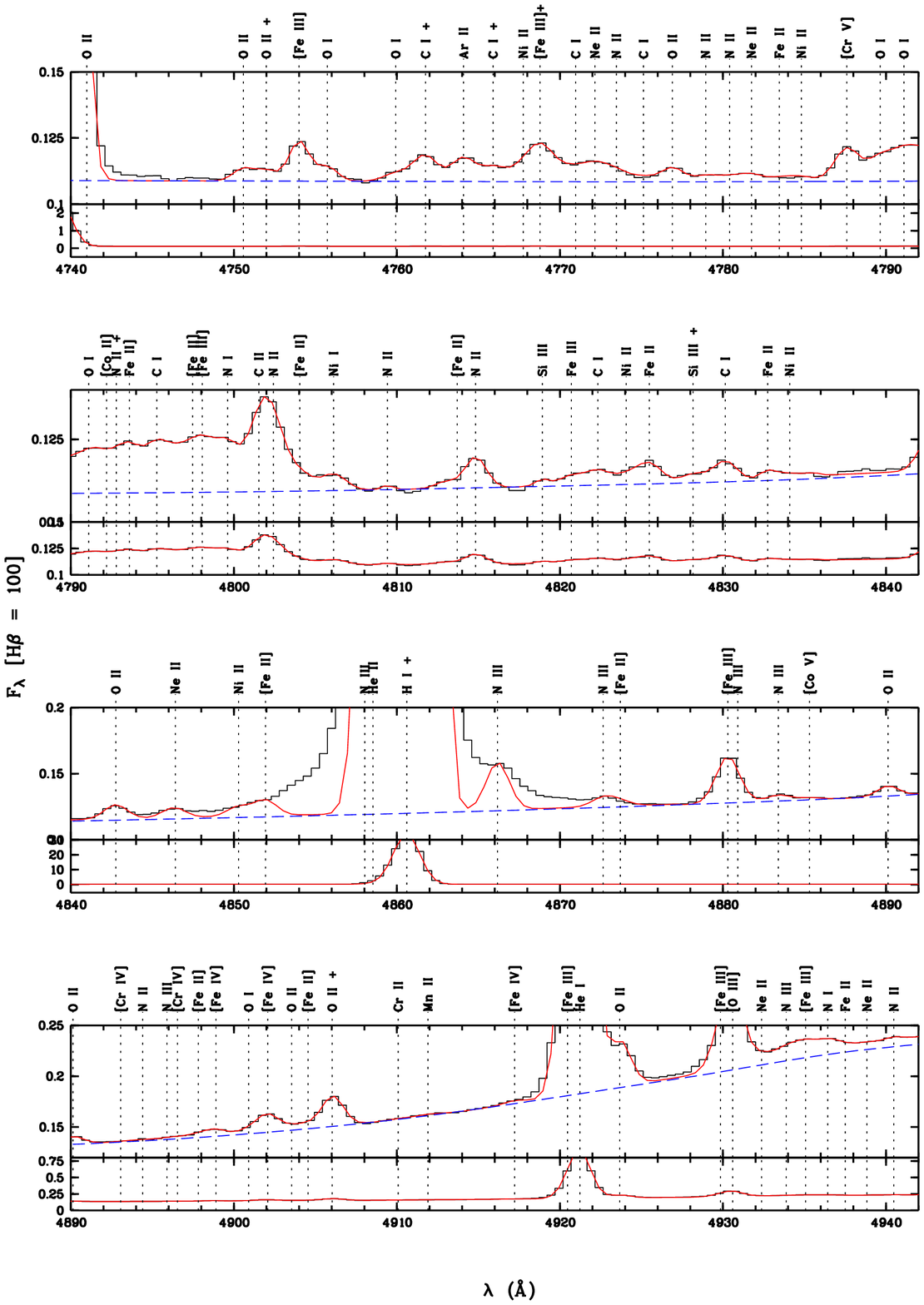,width=15cm,angle=0}
 \caption{Continued.}
 \label{spectra_plot}
\end{center}
\end{figure*}

\addtocounter{figure}{-1}
\begin{figure*}
\begin{center}
 \epsfig{file=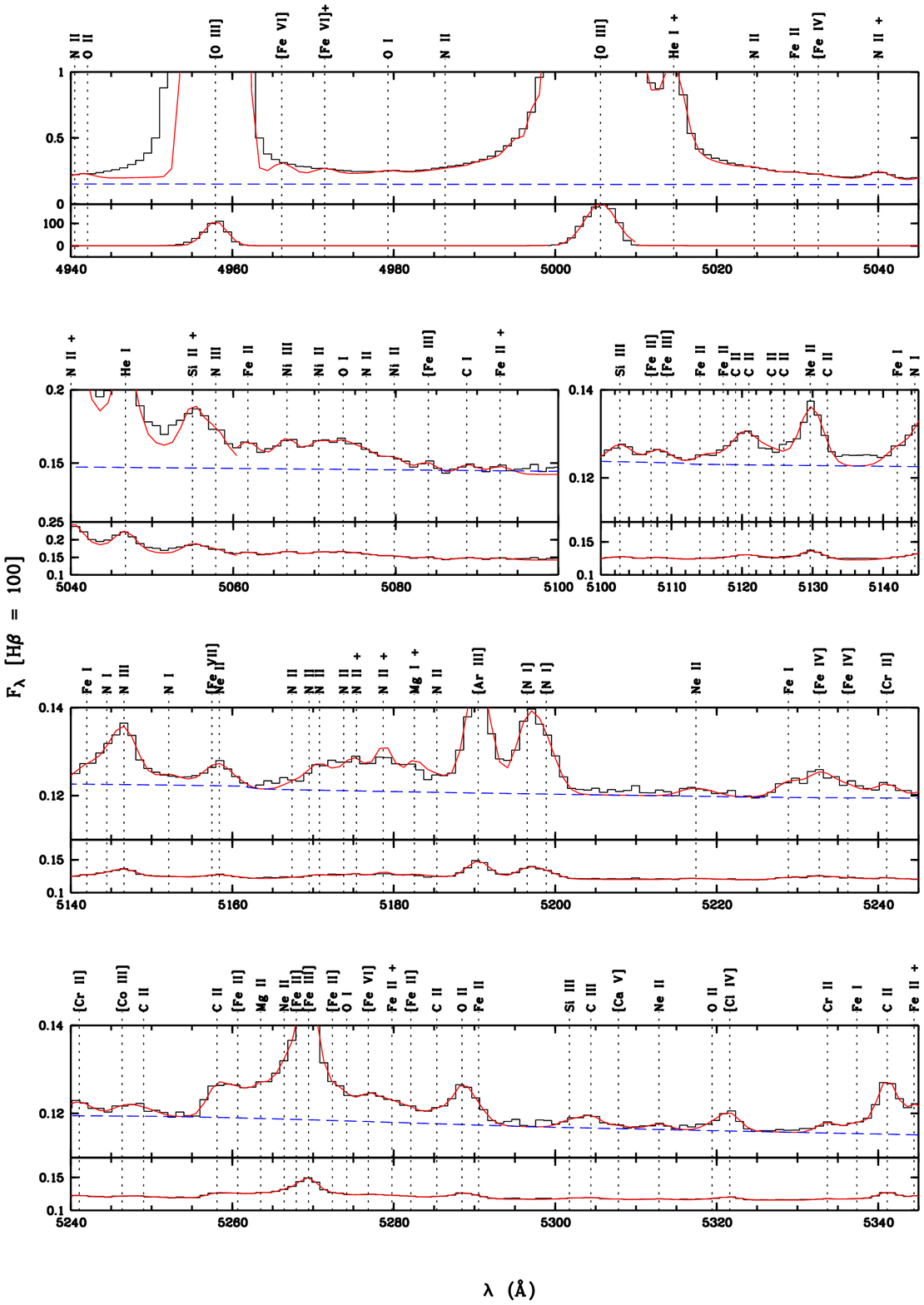,width=15cm,angle=0}
 \caption{Continued.}
 \label{spectra_plot}
\end{center}
\end{figure*}

\addtocounter{figure}{-1}
\begin{figure*}
\begin{center}
 \epsfig{file=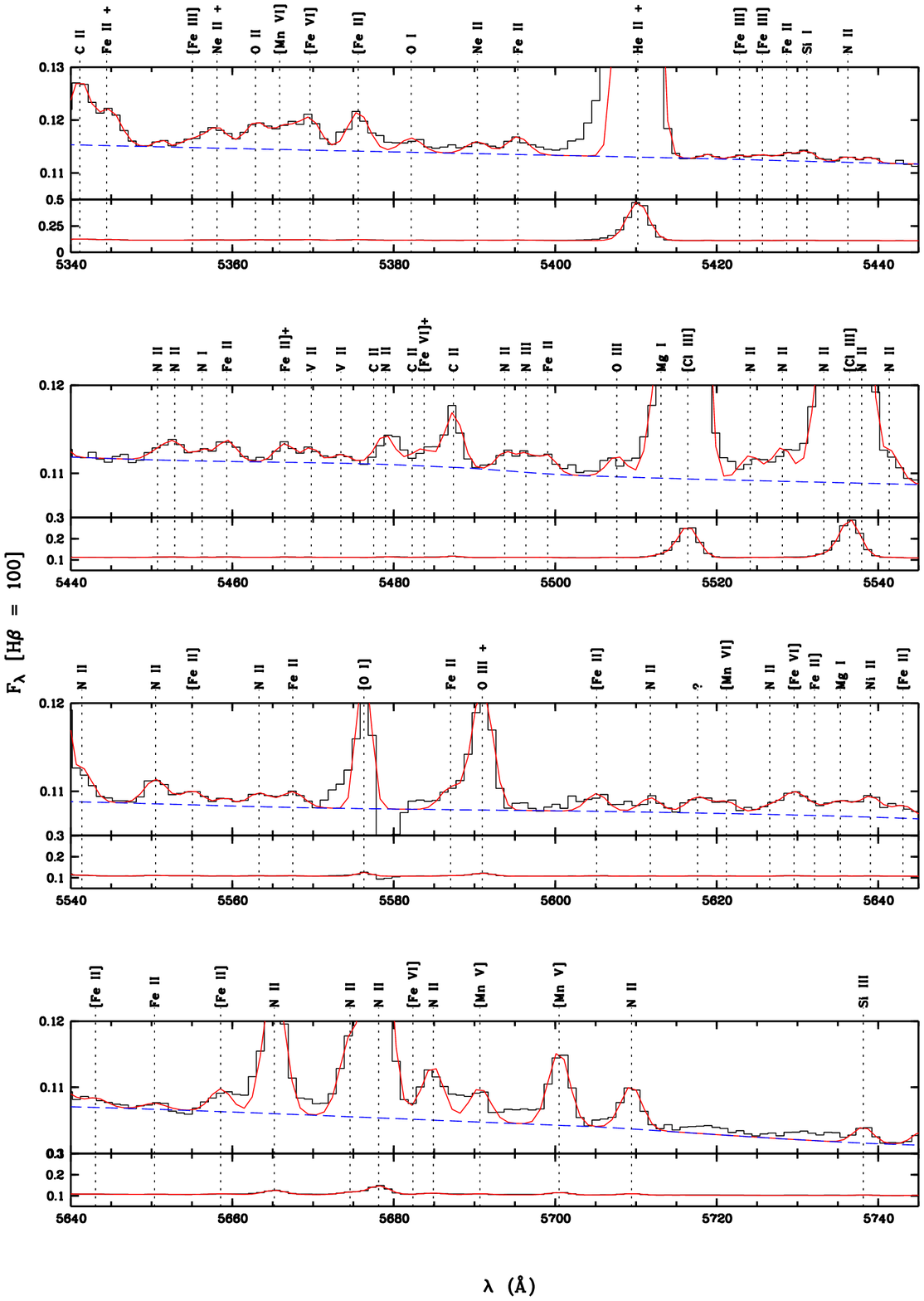,width=15cm,angle=0}
 \caption{Continued.}
 \label{spectra_plot}
\end{center}
\end{figure*}

\addtocounter{figure}{-1}
\begin{figure*}
\begin{center}
 \epsfig{file=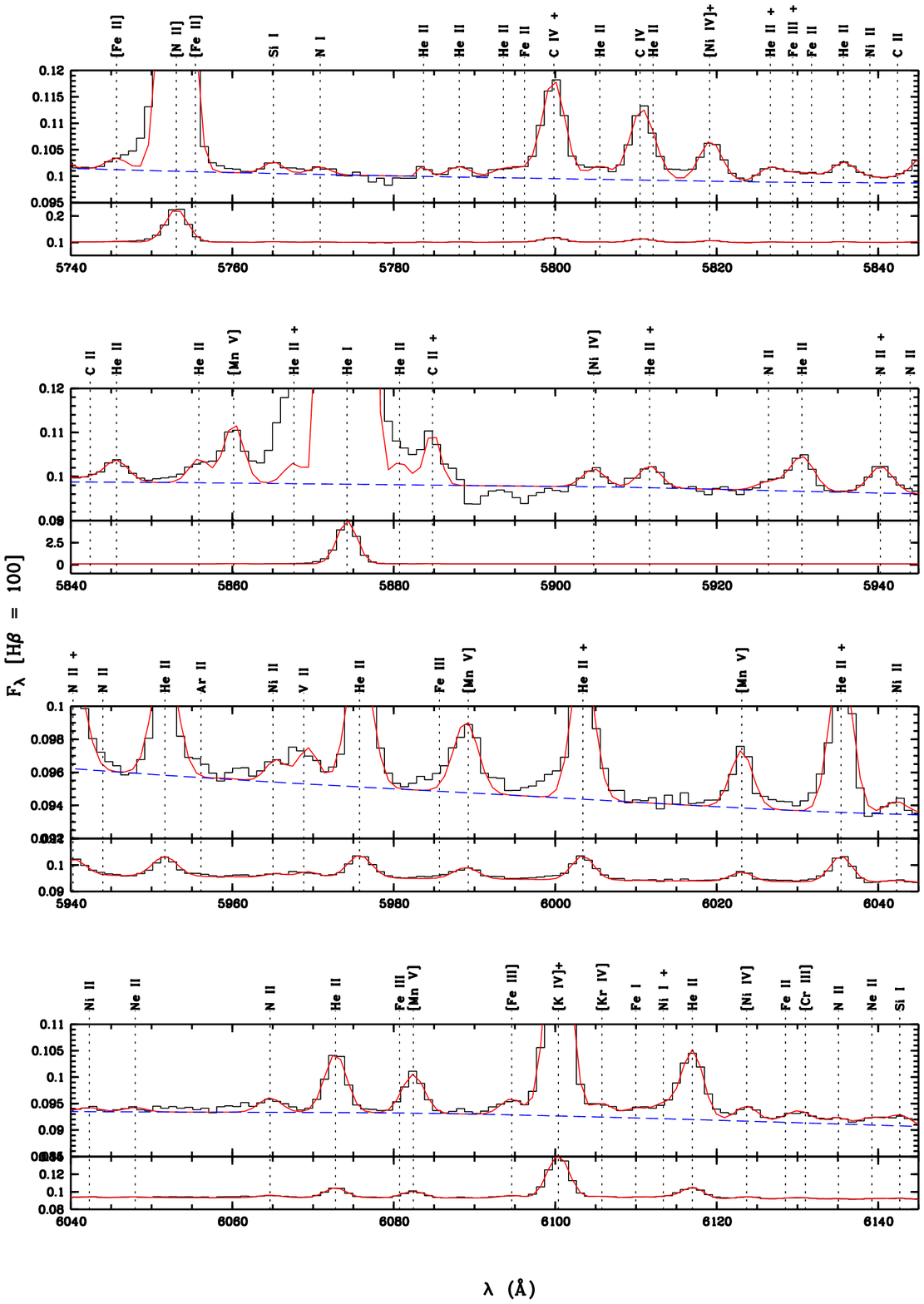,width=15cm,angle=0}
 \caption{Continued.}
 \label{spectra_plot}
\end{center}
\end{figure*}

\addtocounter{figure}{-1}
\begin{figure*}
\begin{center}
 \epsfig{file=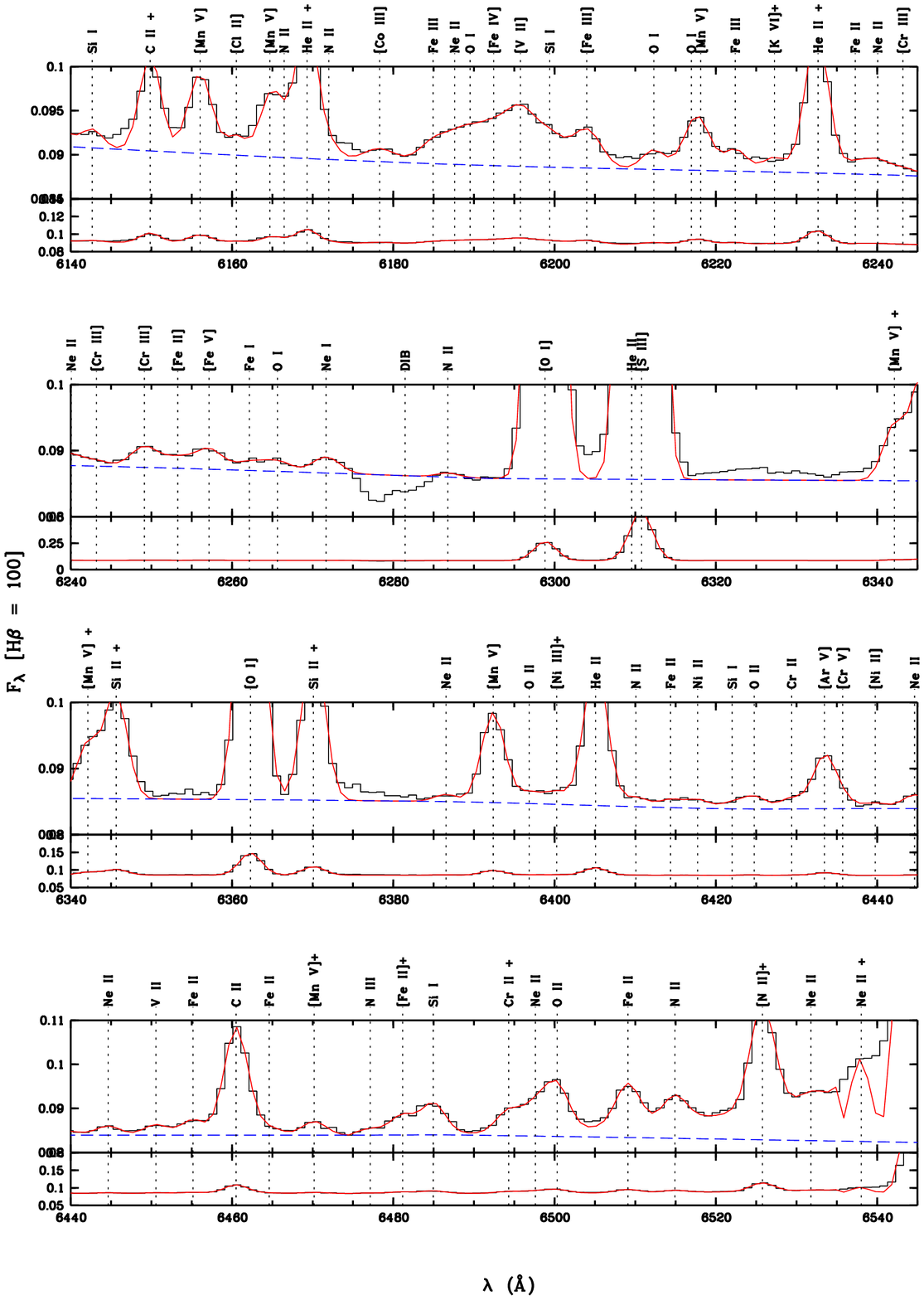,width=15cm,angle=0}
 \caption{Continued.}
 \label{spectra_plot}
\end{center}
\end{figure*}

\addtocounter{figure}{-1}
\begin{figure*}
\begin{center}
 \epsfig{file=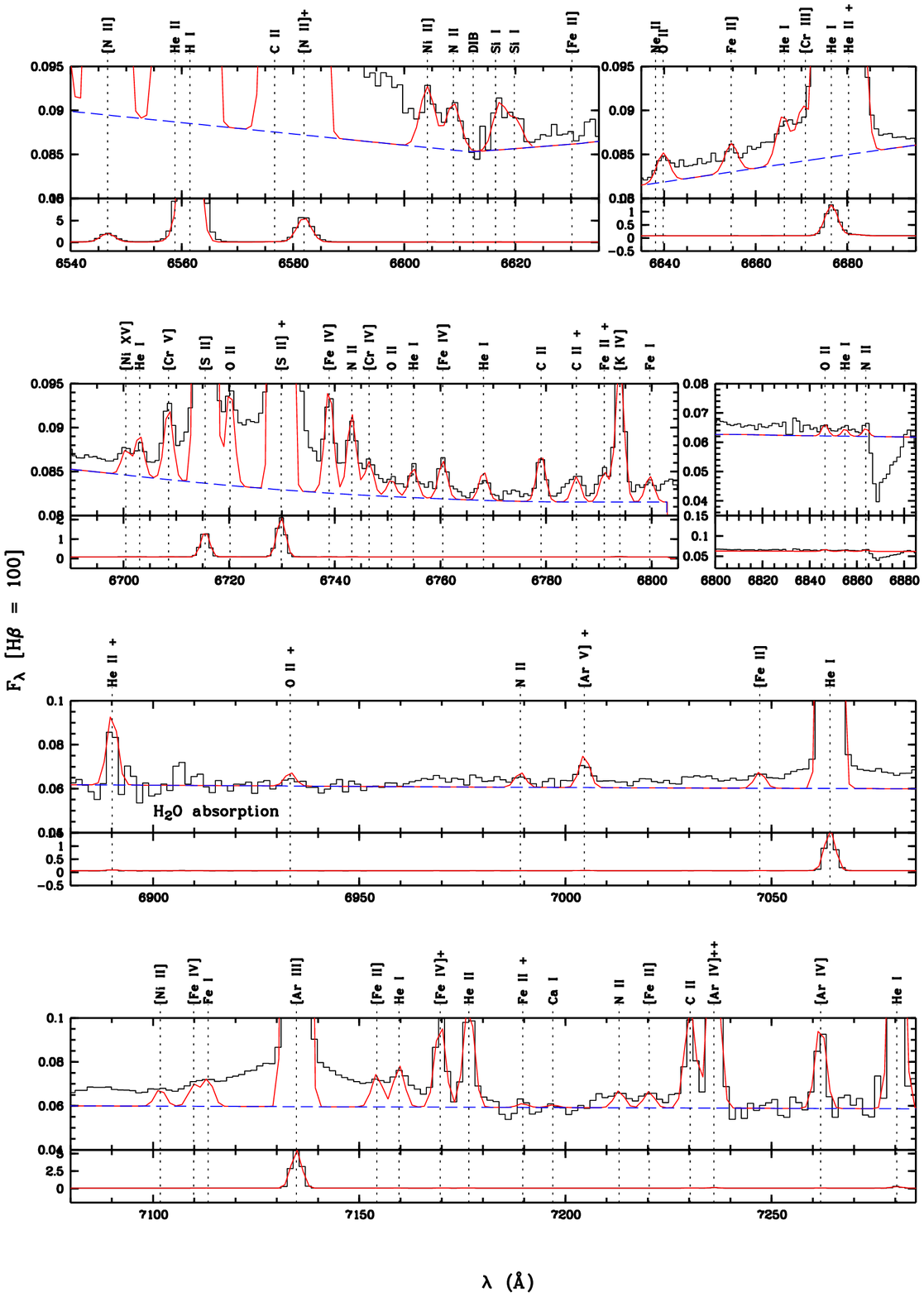,width=15cm,angle=0}
 \caption{Continued.}
 \label{spectra_plot}
\end{center}
\end{figure*}

\addtocounter{figure}{-1}
\begin{figure*}
\begin{center}
 \epsfig{file=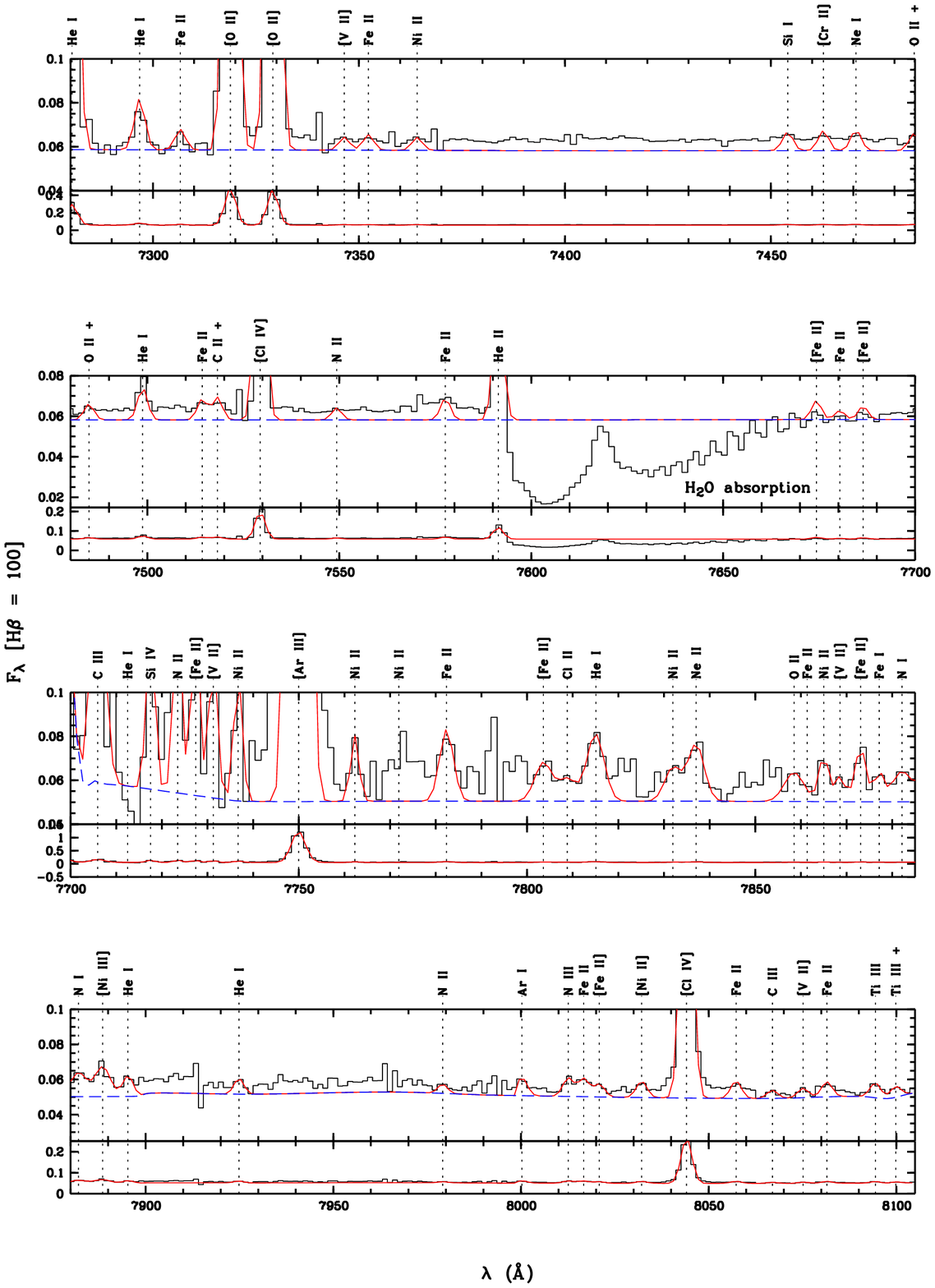,width=15cm,angle=0}
 \caption{Continued.}
 \label{spectra_plot}
\end{center}
\end{figure*}

\addtocounter{figure}{-1}
\begin{figure*}
\begin{center}
 \epsfig{file=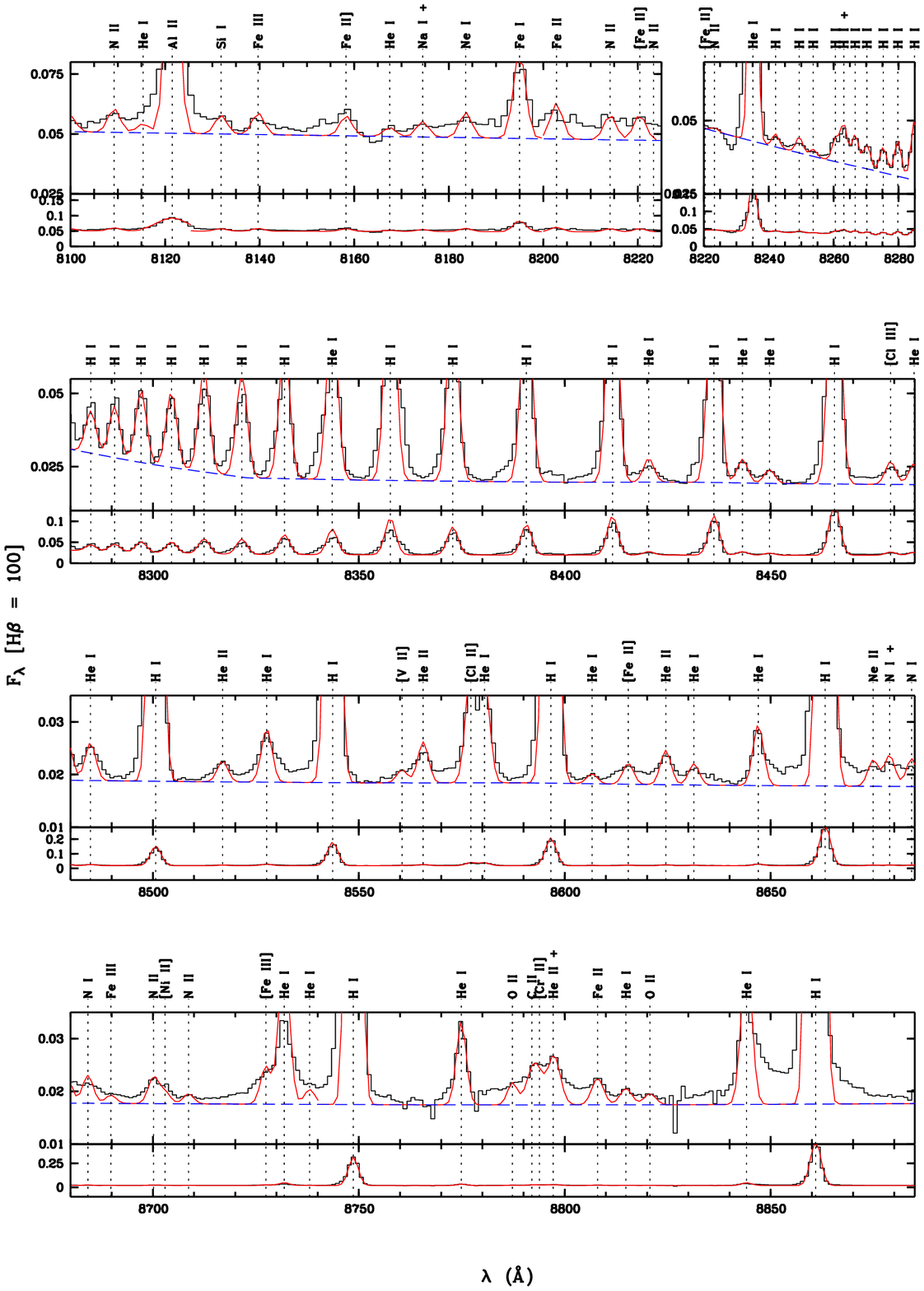,width=15cm,angle=0}
 \caption{Continued.}
 \label{spectra_plot}
\end{center}
\end{figure*}

\addtocounter{figure}{-1}
\begin{figure*}
\begin{center}
 \epsfig{file=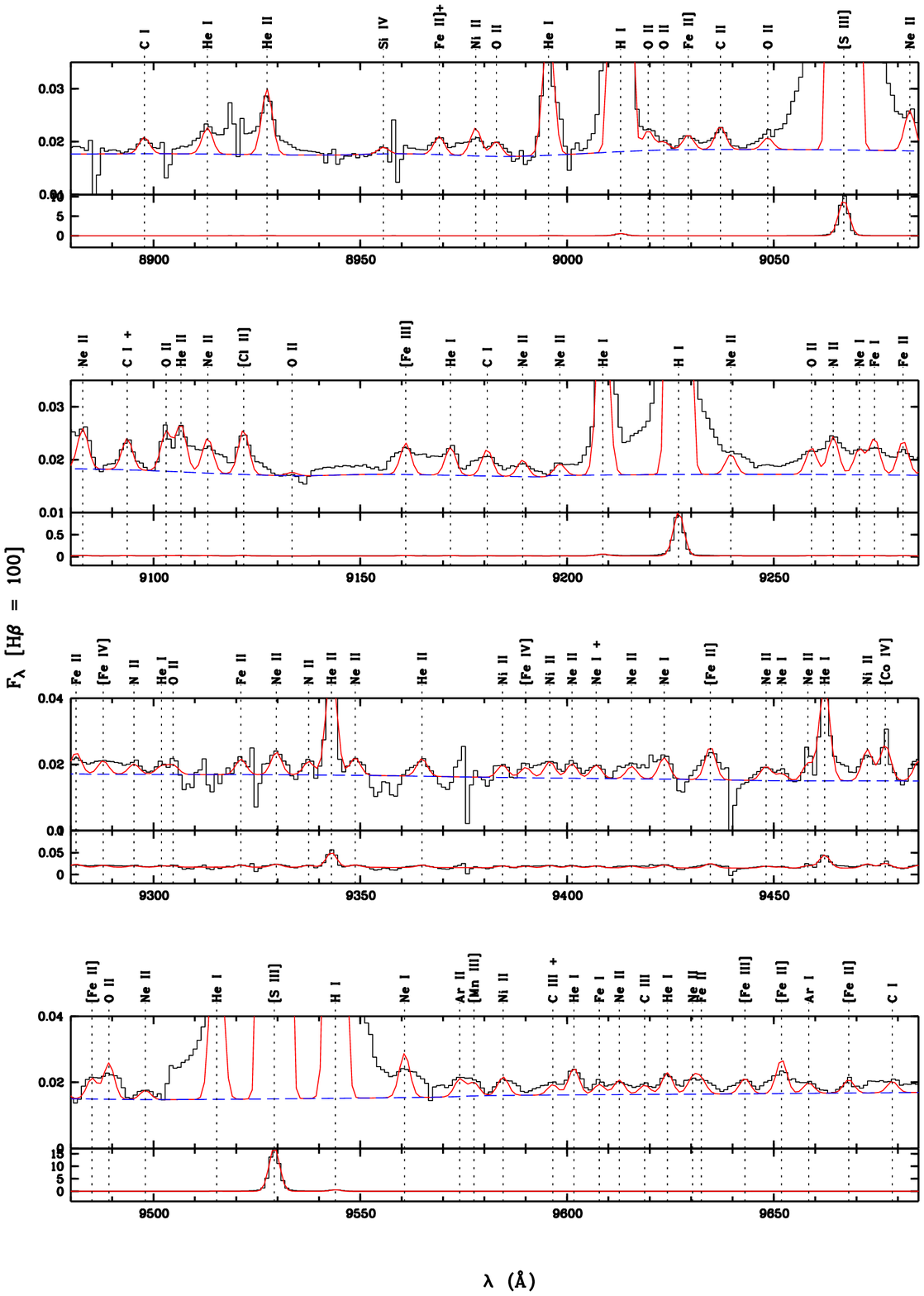,width=15cm,angle=0}
 \caption{Continued.}
 \label{spectra_plot}
\end{center}
\end{figure*}

\addtocounter{figure}{-1}
\begin{figure*}
\begin{center}
 \epsfig{file=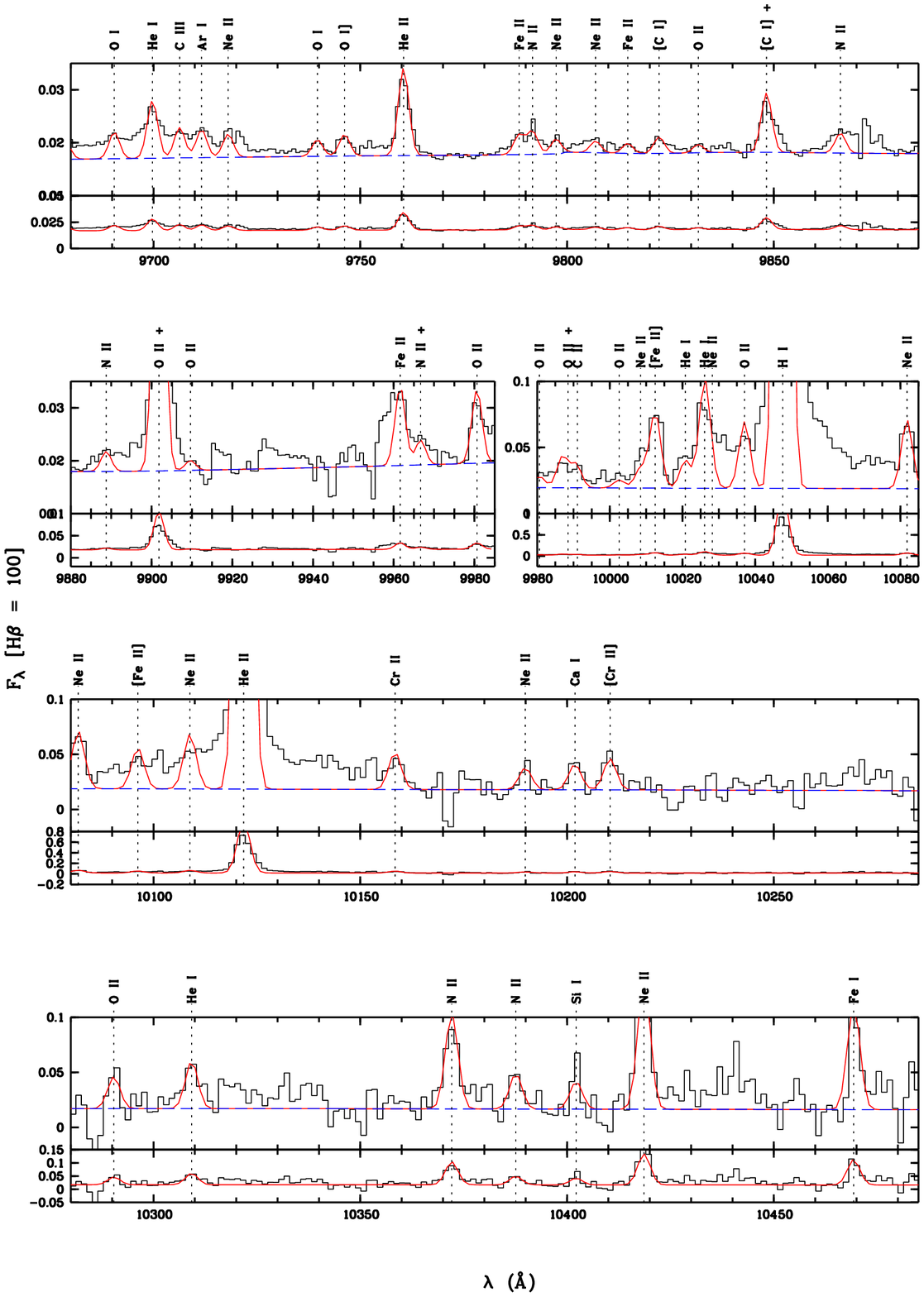,width=15cm,angle=0}
 \caption{Continued.}
 \label{spectra_plot}
\end{center}
\end{figure*}

\clearpage
\addtocounter{figure}{-1}
\begin{figure*}
\begin{center}
 \epsfig{file=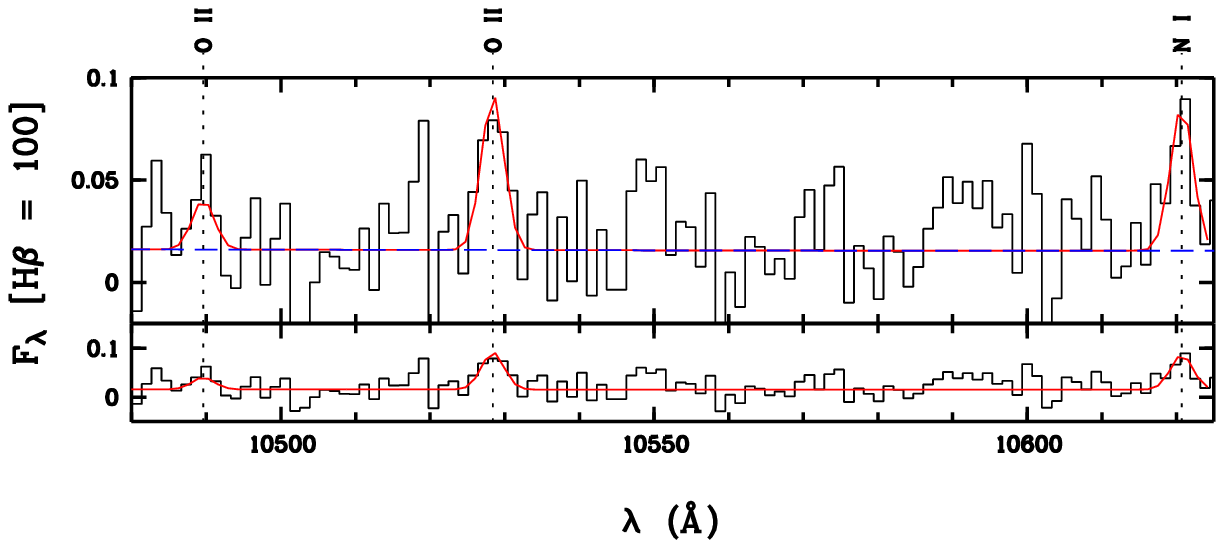,width=8cm,angle=0}
 \caption{Continued.}
 \label{spectra_plot}
\end{center}
\end{figure*}

%

\clearpage
\onecolumn

\begin{longtable}{cccllclllrrr}
\caption{\label{linelist} Emission line list of NGC\,7009. The line fluxes are normalized such that H$\beta$ = 100.}\\
\cline{1-12}\\
$\lambda_\mathrm{obs}$(\AA) & $F(\lambda)$ & $I(\lambda)$ &  & ID & $\lambda_\mathrm{lab}$(\AA) & Mul. & Lower Term & Upper Term & $g_{1}$ & $g_{2}$ & Error (\%)\\
\cline{1-12}\\
\endfirsthead
\caption{Continued.}\\
\cline{1-12}\\
$\lambda_\mathrm{obs}$(\AA) & $F(\lambda)$ & $I(\lambda)$ &  & ID & $\lambda_\mathrm{lab}$(\AA) & Mul. & Lower Term & Upper Term & $g_{1}$ & $g_{2}$ & Error (\%)\\
\cline{1-12}\\
\endhead
\hline
\endfoot
 3041.49   &    0.2480  &    0.3170     &     &    Si II    &  3042.19  &    V14    &   3d' 4D*     &   4f' 4D      &  8 &  8 &       :\\
           &            &           & *\footnote{The asterisk `` * " in front of an identification denotes the line is blended with an adjacent feature, or it is an alternative identification given by {\sc emili}.} &    Si II    &  3042.18  &    V14    &   3d' 4D*     &   4f' 4D      &  4 &  4 &        \\
 3043.75   &    1.3100  &    1.6737     &     &    Ne III   &  3044.24  &           &   3s  3D*     &   3p  3D      &  3 &  5 &    5.75\\
           &            &               &   * &    Ne II    &  3045.56  &    V8     &   3p  4P*     &   3d  4D      &  2 &  2 &        \\
 3045.50   &    3.7120  &    4.7410     &     &    O III    &  3047.12  &    V4     &   3s  3P*     &   3p  3P      &  5 &  5 &    3.03\\
 3046.82   &    0.0950  &    0.1213     &     &    Ne II    &  3047.56  &    V8     &   3p  4P*     &   3d  4D      &  4 &  6 &   86.32\\
 3058.27   &    1.0840  &    1.3814     &     &    O III    &  3059.30  &    V4     &   3s  3P*     &   3p  3P      &  5 &  3 &    7.56\\
           &            &               &   * &    C V      &  3057.50  &           &   3s  3S      &   3p  3P*     &  3 &  5 &        \\
\end{longtable}

\section*{acknowledgements}
X. Fang and X.-W. Liu thank P.~J. Storey for making the O~{\sc ii} effective 
recombination coefficients available prior to publication.

\end{document}